\documentclass{PoS}

\title{The azimuth quadrupole in nuclear collisions}

\ShortTitle{Azimuth Quadrupole}

\author{\speaker{David Kettler for the STAR Collaboration}\\
        University of Washington\\
        E-mail: \email{dkettler@u.washington.edu}}

\abstract{

We present measurements of both $p_t$-integral and $p_t$-differential
azimuth quadrupole components of two-particle correlations on azimuth ($\phi$)
and pseudorapidity ($\eta$) for unidentified hadrons in Au-Au collisions at
$\sqrt{s_{NN}}=62$ and 200 GeV.  The azimuth quadrupole component is
distinguished from $\eta$-localized same-side correlations by taking advantage
of the $\eta$ dependence.  The quadrupole component is
related to conventional $v_2$ measures.  Both $p_t$-integral and
$p_t$-differential results are presented as functions of Au-Au centrality.  We
observe simple universal energy and centrality trends for the
$p_t$-integral quadrupole component.  $p_t$-differential results are
constructed using $v_{2}^{2}$ marginal distributions on $p_t$.  These results can
be transformed to reveal quadrupole $p_t$ spectra that are nearly independent
of centrality.  A parametrization of the $p_t$-differential quadrupole
shows a simple $p_t$ dependence that can be factorized from the centrality
and collision energy dependence above 0.75~GeV/c.

}

\FullConference{Workshop on Critical Examination of RHIC Paradigms - CERP2010\\
		April 14-17, 2010\\
		Austin Texas USA}

\begin{document}


\section{Introduction}

One of the major physics results of heavy ion collisions at RHIC energies has been the observation of a large azimuth anisotropy \cite{starreview}.  This phenomenon is typically described in terms of hydrodynamics in which there is a pressure-driven expansion of the collision region in non-central collisions \cite{hydrointro}.
Hydrodynamics describes both radial flow (a collective transverse velocity of particles produced in the collision) and elliptic flow (the azimuth anisotropy produced from the asymmetric shape of the original collision system).

Elliptic flow is typically thought to correspond to the measured quantity $v_2$, defined by the second Fourier component of the distribution of particles on azimuth with respect to the angle of the reaction place of the colliding nuclei \cite{basicv2}.  This definition works for theoretical calculations in which the reaction plane is known.  However, in experimental measurements the reaction plane can only be estimated by particles in the event, so correlations must be used.


Two-particle correlations could reliably measure the phenomenon of elliptic flow if there were no other physical contributions to the correlation signal \cite{azstruct}.  In practice there are several sources of correlations such as (mini)jets, resonances, and HBT effects.  There are two major approaches for reducing so-called ``nonflow'' effects.   One is to use the pseudorapidity separation between particles and the other is to make a many-particle measurement.

In this analysis we distinguish the azimuth quadrupole from other sources of correlation geometrically by using the shape of the correlations on pseudorapidity as well as azimuth.  Full two-dimensional histograms of the two-particle correlation space are constructed and fitted with model functions to distinguish different contributions.

\section{Angular Correlations}
\label{correlations}

Minimum-bias angular correlations are constructed by considering all possible pairs of particles in an event (minus self-pairs).  The primary measurement variables of the particles we observe are their azimuth angle $\phi$, pseudorapidity $\eta$, and transverse momentum $p_t$.  In two-particle correlations these three variables for each particle define a six-dimensional space $(\phi_1 , \eta_1 , p_{t1} , \phi_2 , \eta_2 , p_{t2})$.  This space is difficult to work with, so cuts and projections are used to simplify the analysis.

\begin{figure*}
  \hspace{90pt}
\resizebox{.50\textwidth}{!}{
  \includegraphics{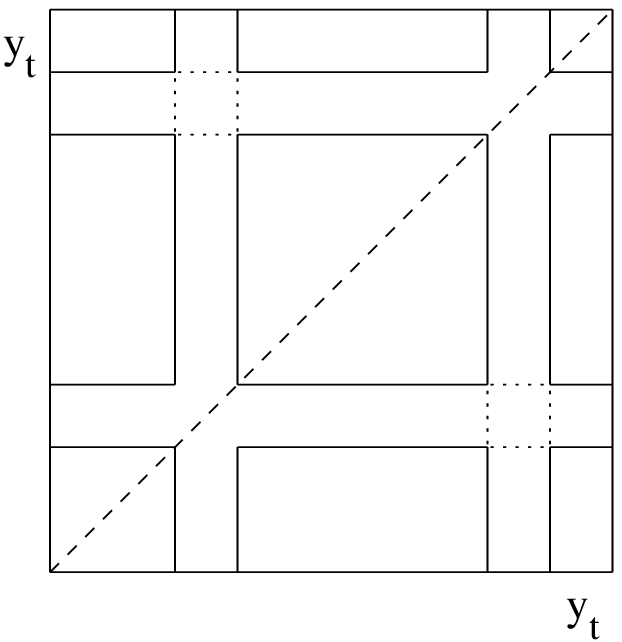}
  \resizebox{.42\textwidth}{!}{\includegraphics{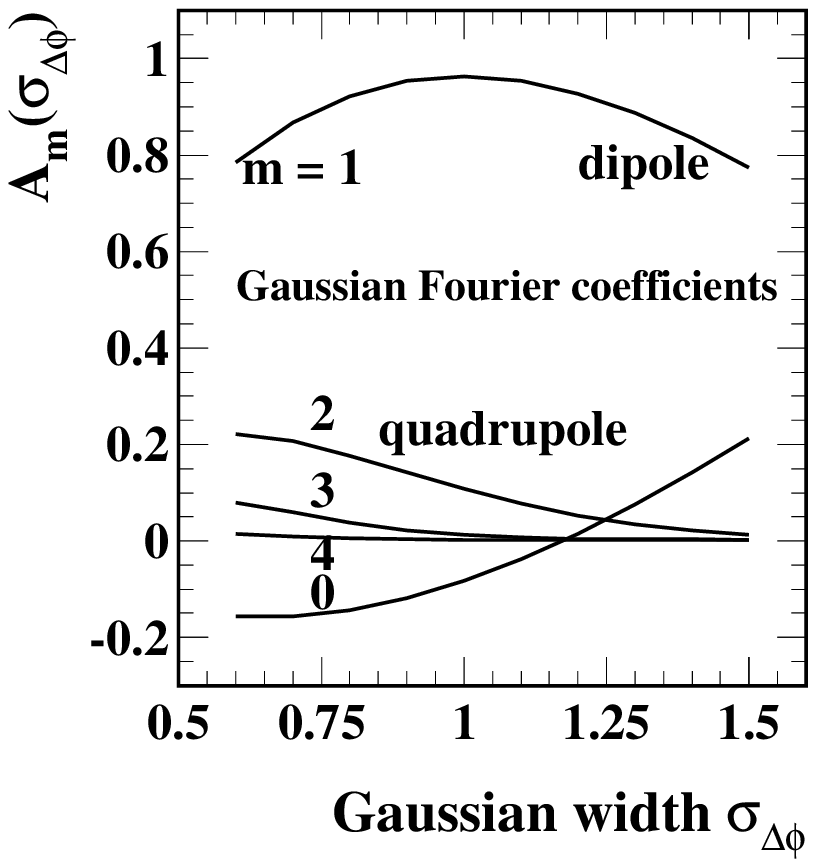}}
}
\caption{\label{fig:diagrams} Left Panel: Two example of marginal cuts in $y_t \times y_t$ space.  Right Panel: The Fourier decomposition of a periodic array of Gaussians as a function of the Gaussian width.
}
\end{figure*}

In this analysis the first cut selects a momentum range from the $( p_{t1} , p_{t2} )$ space.  One possible choice is to accept the entire momentum range, which yields what we refer to as a $p_t$-integrated correlation.  However, if we want to study the $p_t$ dependence of correlations there are several choices, since we are dealing with a two-dimensional $p_t$ space.  The momentum cuts we have chosen are marginal distributions in which we restrict the momentum range of one of the particles and allow the other particle to have any momentum value.  Because of the inherent diagonal symmetry in the $( p_{t1} , p_{t2} )$ space this produces cross-shaped cuts as seen in Fig.~\ref{fig:diagrams} (left panel).

It is convenient to use transverse rapidity $y_t$ instead of transverse momentum $p_t$.  $y_t$ is a relativistic momentum variable defined as
\begin{eqnarray}
y_t = \ln \left\{ (p_t + m_t ) / m_0 \right\}.
\label{eqn:yt}
\end{eqnarray}
$y_t$ is analogous to the usual longitudinal rapidity variable but is defined in the transverse direction with $y_z \approx 0$.  It is useful for studying a large range of momentum values.  The definition of $y_t$ requires a mass, but in practice we usually analyze unidentified particles so the pion mass is assumed.  When used in this fashion $y_t$ is a logarithmic transformation of $p_t$ that is very close to $\ln ( p_t )$ but has a well-defined zero.

The correlation measure is constructed from pair densities of sibling pairs (pairs of particles taken from the same event) and mixed pairs (pairs of particles taken from similar but distinct events).  Mixed pairs serve as a reference that can remove detector artifacts in the measure $\Delta \rho / \rho_{\textrm{ref}} \equiv ( \rho_{\textrm{sib}} - \rho_{\textrm{ref}} ) / \rho_{\textrm{ref}}$, with the mixed-pair reference normalized to the total number of the sibling pairs.  This is a per-pair measure.  At times a per-particle measure such as $\Delta \rho / \sqrt{\rho_{\textrm{ref}}}$ is more useful since a per-pair measure includes a trivial $1 / n_{ch}$ trend.

This approach can be contrasted with the ZYAM (zero yield at minimum) method which is applicable only for narrow, well-separated peaks \cite{ZYAM}.  The difference is in the normalization of mixed pairs to sibling pairs.  The ZYAM method imposes the criterion that the minimum of the correlation histogram should have value zero and adjusts the normalization in order to achieve that.  But if the physics signal contains distinct but overlapping peaks then the true correlation value of the minimum of the histogram should be nonzero.  This normalization also affects the amplitudes of the peaks.

This analysis is based on 14.5 million Au-Au collisions with center-of-mass energy 200~GeV and 6.7 million collisions at 62~GeV.  Tracks are observed in the STAR TPC with a minimum $p_t$ of $0.15 \textrm{ GeV}/c$ and no upper $p_t$ cut.  The TPC has $2 \pi$ coverage in azimuth and this analysis uses tracks in $-1 < \eta < 1$.
The analysis also divides the minimum-bias events into eleven different centrality classes as defined in Ref.~\cite{centralities}, with nine $\sim 10\%$ bins from 100\% to 10\% and the most-central 10\% divided into two 5\% bins.


It is useful to compare these correlations to the standard event-plane method for measuring $v_2$ \cite{basicv2},  defined by
\begin{eqnarray}
v_m \{ EP \} = \frac{v_{m}^{obs}}{\left< \cos ( m ( \psi_m - \psi_r ) ) \right>} ,
\label{eqn:v2epDef}
\end{eqnarray}
where $v_{m}^{obs} = \left< \cos ( m ( \phi - \psi_m ) ) \right>$, $\psi_r$ is the true reaction plane angle, $\psi_m$ is the estimated event-plane angle, and $\phi$ is the azimuth angle of a particle in the event.  The denominator of this expression is known as the event-plane resolution.  $\psi_m$ is measured as the direction of the corresponding $Q$ vector
\begin{eqnarray}
\vec{Q}_m = \sum_{i=1}^{n} \vec{u} ( m \phi_i ) \equiv Q_m \vec{u} ( m \psi_m ) .
\label{eqn:QDef}
\end{eqnarray}
Here $\vec{u}$ are unit vectors and $Q_m$ is the magnitude of the $Q$ vector.  The two-particle correlation density $V_m$ can be defined by the expression
\begin{eqnarray}
V_{m}^{2} = \sum_{i=1}^{n} \sum_{j \neq i}^{n-1} \vec{u} ( m \phi_i ) \vec{u} ( m \phi_j ) = n^2 v_{m}^{2} \{ 2 \} ,
\label{eqn:VmDef}
\end{eqnarray}
where $n$ is the event multiplicity.  The second part of the above expression introduces the relationship to the two-particle cumulant method.
The $V_m$ here are strictly Fourier components of the total two-particle azimuth density and include ``nonflow'' effects that might be better described by non-Fourier terms.

If we use Eq.~\ref{eqn:QDef} in Eq.~\ref{eqn:VmDef} then we arrive at the following relationship:
\begin{eqnarray}
V_{m} = n \left \{ \frac{1}{n} \sum_{i=1}{n} \vec{u} ( m \phi_i ) \right \} \cdot \vec{u} ( m \psi_m ) \frac{Q_m}{V_m} .
\label{eqn:VmMath}
\end{eqnarray}
In Ref.~\cite{azstruct} an expression for the event-plane resolution was derived:
\begin{eqnarray}
\left< \cos [ m ( \psi_m - \psi_r ) ] \right> \approx \sqrt{\frac{n-1}{n}} \frac{V_m}{Q_m} .
\label{eqn:EPRes}
\end{eqnarray}
Inserting this into Eq.~\ref{eqn:VmMath} gives:
\begin{eqnarray}
V_m \approx n \frac{v_{m}^{obs}}{\left< \cos ( m ( \psi_m - \psi_r ) ) \right>} = n v_m \{ EP \} .
\label{eqn:EPRes}
\end{eqnarray}
Thus we find that the event-plane method of measuring $v_2$, while often presented using different language, is essentially a type of two-particle correlation.

\section{$p_t$--integrated Quadrupole}
\label{integrated}


$p_t$-integrated quadrupole results have been presented previously in \cite{quadrupole} but are still important for understanding the $p_t$-differential results.  The systematics they establish provide a basis for fully understanding the correlation system.

The quadrupole component is obtained by doing a free fit of two-dimensional (2D) $(\eta_{\Delta}, \phi_{\Delta})$ histograms.  It has been found that these histograms can be well-described at all centralities by a remarkably simple fit model which works even in proton-proton collisions.  The model function includes: a same-side 2D Gaussian on $(\eta_\Delta , \phi_\Delta )$, and $\eta_\Delta$-independent away-side dipole $\cos (\phi_\Delta - \pi)$, an $\eta_\Delta$-independent quadrupole $\cos (2 \phi_\Delta )$, a $\phi_\Delta$-independent 1D Gaussian on $\eta_\Delta$, a narrow same-side 2D exponential on $(\eta_\Delta , \phi_\Delta )$, and a constant normalization offset.  The model function is expressed as
\begin{eqnarray}
F &=& A_{D} \cos (\phi_\Delta - \pi) + A_{Q} \cos (2 \phi_\Delta ) + A_0 e^{- \frac{1}{2} \left( \frac{\eta_\Delta}{\sigma_0} \right) ^2}
+ A_1 e^{- \frac{1}{2} \left\{ \left( \frac{\phi_\Delta}{\sigma_{\phi_\Delta}} \right) ^2 + \left( \frac{\eta_\Delta}{\sigma_{\eta_\Delta}} \right) ^2 \right\}} \nonumber \\
&+& A_2 e^{- \left\{ \left( \frac{\phi_\Delta}{w_{\phi_\Delta}} \right) ^2 + \left( \frac{\eta_\Delta}{w_{\eta_\Delta}} \right) ^2 \right\} ^{1/2} } + A_3 .
\label{eqn:fitfunc}
\end{eqnarray}

The quadrupole term is
conventionally interpreted as ``elliptic flow'' \cite{azstruct}.  The other important structures in this study are the 2D same-side peak and corresponding away-side ridge, described by a 2D Gaussian and the away-side dipole respectively.  The same-side 2D peak can contain contributions from HBT and resonance decays, but is likely dominated by minijets \cite{mikeQM, fragevolve}.
The sharp 2D exponential peak mainly describes electron pair production and is not of interest in this study.  The 1D Gaussian on $\eta_\Delta$ is only present in more-peripheral collisions and is believed to be related to participant nucleon fragmentation.  Given that it has no $\phi_\Delta$ dependence it is orthogonal to the quadrupole term which is only $\phi_\Delta$-dependent.

A more complete description of the away-side structure would be with a periodic array of Gaussian peaks centered at $\pi$, $3 \pi$, etc.  The dipole description applies in the limit where the widths of the Gaussian peaks become large.  Fig.~\ref{fig:diagrams} (right panel) shows the calculated Fourier coefficients of such an array of Gaussians as a function of the Gaussian width.  In the $p_t$-integrated results the dipole appears to be a good description for all centralities.

\begin{figure*}
\resizebox{1.00\textwidth}{!}{
  \includegraphics{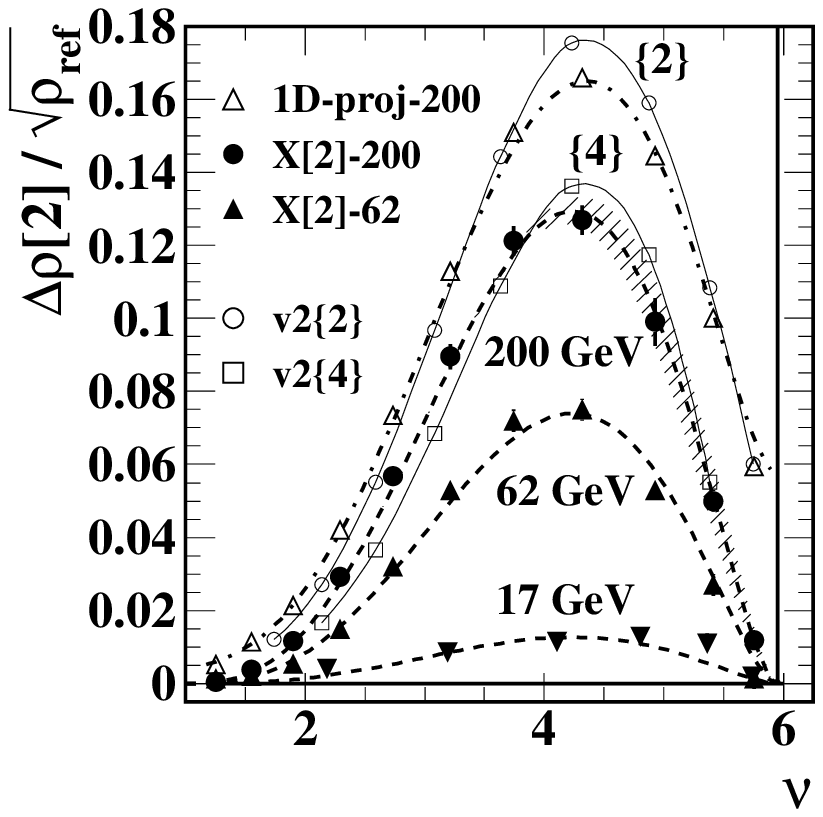}
  \includegraphics{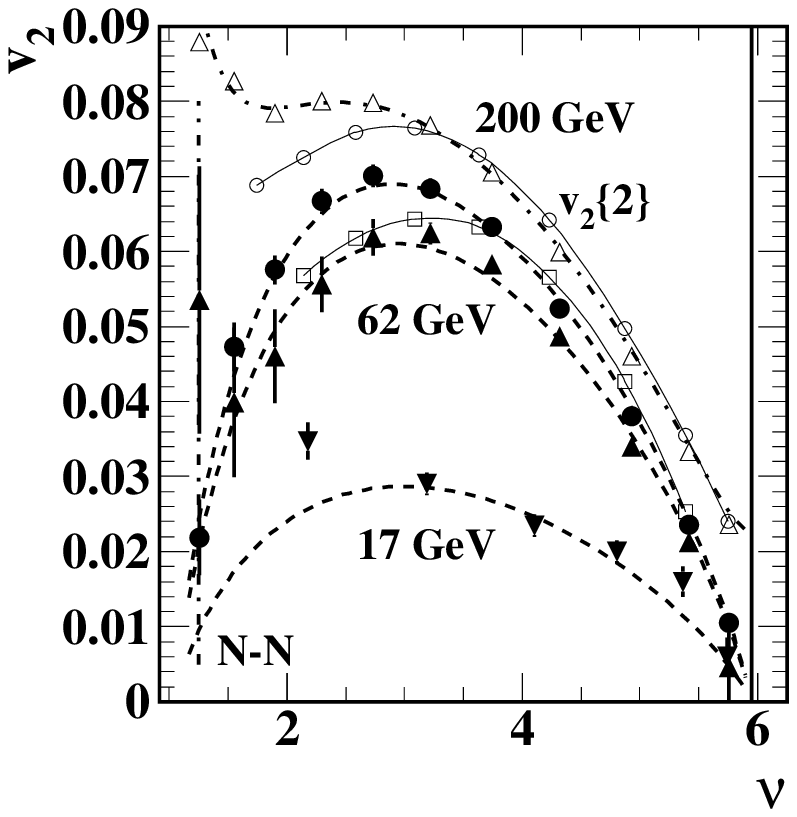}
  \includegraphics{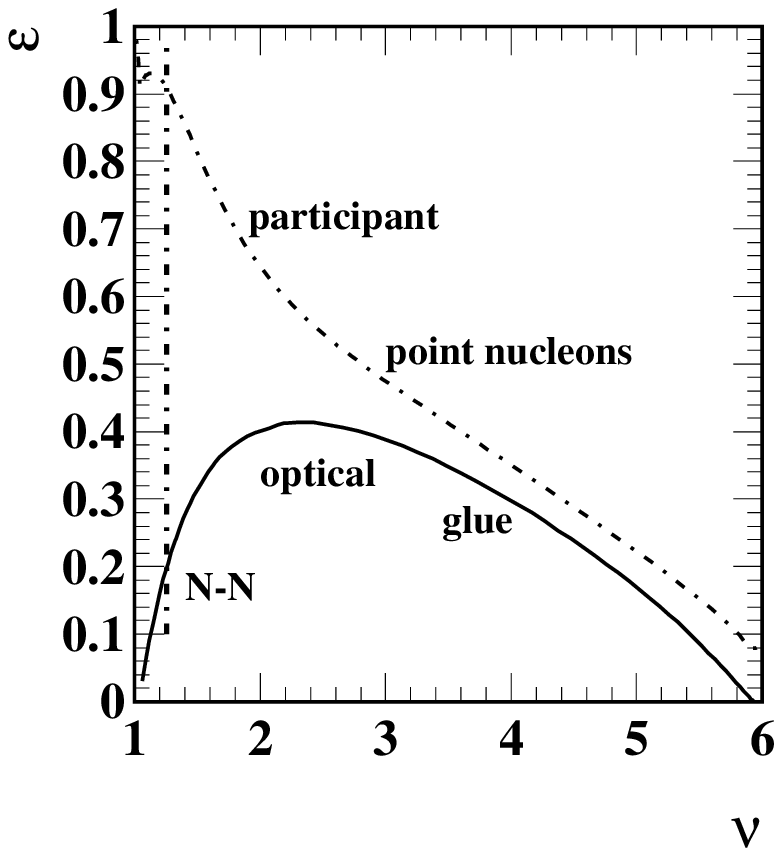}
  \includegraphics{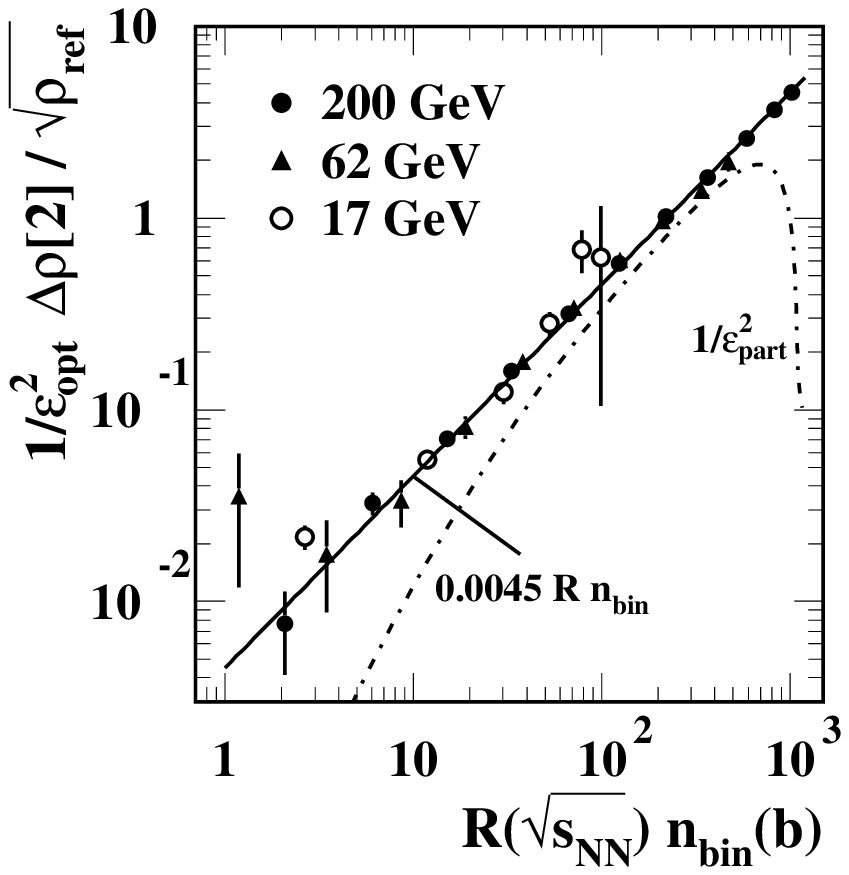}
}
\caption{\label{fig:ptint} First Two Panels: $p_t$-integrated quadrupole results in terms of $\Delta \rho / \sqrt{\rho_{\textrm{ref}}}$ and $v_2$.  The solid circles and solid triangles are quadrupole results from 2D fitting at 200 and 62 GeV respectively.  The open circles and open squares are $v_2 \{ 2 \}$ and $v_2 \{ 4 \}$ results for 200 GeV collisions \cite{v24}.  The upside-down solid triangles are 17 GeV results from NA49 \cite{na49}.  The open triangles are from projecting the 2D correlations to just $\phi_\Delta$ and fitting with only Fourier components.  The dashed curves are from our parametrization of the data.  Third Panel: Calculated optical and participant eccentricities as a function of centrality.  Right Panel: Energy and centrality dependence of the quadrupole using the optical eccentricity for 200 and 62 GeV data from this study and 17 GeV results from NA49 \cite{na49}.
}
\end{figure*}

The quadrupole component in $\Delta \rho / \sqrt{\rho_{\textrm{ref}}}$ is related to the usual $v_2$ measure by \cite{azstruct}
\begin{eqnarray}
A_{Q} = 2 \frac{\Delta \rho [2]}{\sqrt{\rho_{\textrm{ref}}}} \equiv 2 \rho_0 v_{2}^{2} (b).
\end{eqnarray}
Extracted quadrupole parameters are shown in Fig.~\ref{fig:ptint} (first two panels) in terms of both $\Delta \rho / \sqrt{\rho_{\textrm{ref}}}$ and $v_2$ amplitudes and compared to published STAR $v_2 \{ 2 \}$ and $v_2 \{ 4 \}$ \cite{v24} data and 17~GeV event-plane results from NA49 \cite{na49}.

In order to better understand the energy and centrality dependence of the azimuth quadrupole it is useful to understand the geometry of the initial collision system which is usually described in terms of an eccentricity \cite{ecc}
\begin{eqnarray}
\epsilon = \frac{\left< y \right>^{2} - \left< x \right>^{2}}{\left< y \right>^{2} + \left< x \right>^{2}},
\label{eqn:eccDef}
\end{eqnarray}
where $x$ and $y$ are coordinates in the plane perpendicular to the beam axis with $x$ being in the reaction plane.  Eccentricity is not directly measurable.  It is typically estimated based on the centrality of the collision using a Glauber model.  The optical model of the eccentricity used here is based on a continuous transverse density profile with a Woods-Saxon potential as in
Ref.~\cite{jaccoop}.

In Ref.~\cite{quadrupole} we show a parametrization of the optical eccentricity for RHIC energies
\begin{eqnarray}
\epsilon_{opt} ( n_{bin} ) = \frac{1}{5.68} \log_{10} \left ( \frac{3 n_{bin}}{2} \right )^{0.96} \cdot \log_{10} \left( \frac{1136}{n_{bin}} \right)^{0.81}.
\label{eqn:OpticalEcc}
\end{eqnarray}
This is plotted as the solid curve in Fig.~\ref{fig:ptint} (third panel).  The dash-dotted curve shows an example of a participant nucleon calculation which is preferred by some to describe conjectured ``flow fluctuations'' \cite{part}.  Participant nucleon models give larger eccentricity values at all centralities but especially in the very peripheral and central cases.

In the quadrupole data in Fig.~\ref{fig:ptint} (first panel) we note that the data at different energies can be described by a common shape simply by varying the amplitude.  The energy dependence is found to be proportional to $\log ( \sqrt{s_{NN}} / 13 \textrm{ GeV} )$.  We then introduce the energy scaling factor
\begin{eqnarray}
R(\sqrt{s_{NN}}) \equiv \log ( \sqrt{s_{NN}} / 13 \textrm{ GeV} ) / \log (200/13) .
\label{eqn:energyScale}
\end{eqnarray}
The complete set of quadrupole data can then be described by 
\begin{eqnarray}
\rho_0 ( b ) v_{2}^{2} \{ 2D \} ( b ) = 0.0045 R ( \sqrt{s_{NN}} ) \epsilon^2 ( b ) n_{bin} ( b ) .
\label{eqn:EnCentDep}
\end{eqnarray}
In Fig.~\ref{fig:ptint} (fourth panel) we confirm this by observing the linear trend when plotting $(1 / \epsilon^2 ) \Delta \rho [2] / \sqrt{\rho_{ref}}$ vs. $R(\sqrt{s_{NN}}) n_{bin} (b)$.  The parametrization of the quadrupole is very useful because it factorizes the energy and centrality dependence.

We can express the $p_t$-integrated $v_2$ at centrality $b$ in terms of the $p_t$-differential $v_2$ by
\begin{eqnarray}
v_{2} (b) = \frac{1}{\rho_0 (b)} \int dp_t p_t \rho_0 (p_t , b) v_2 (p_t , b),
\label{eqn:v2pt}
\end{eqnarray}
where $\rho_0 (p_t , b)$ is the single-particle $p_t$-spectrum \cite{tcspectra}.  Because the spectrum falls off exponentially at larger $p_t$, $p_t$-integrated $v_2$ numbers are heavily weighted toward lower-$p_t$ particles and
have virtually no sensitivity to $v_2$ above about $0.5$~GeV/c.
$p_t$-integrated $v_2$ results should not be used to conclude anything about the behavior of $v_2$ at higher $p_t$.

\section{$p_t$--differential Quadrupole}


The $y_t$ dependence of angular correlations is studied by making cuts on $y_t$ and examining the 2D histograms in restricted $y_t$ intervals.  For two-particle correlations it is possible to restrict the transverse momentum of each particle independently, so the $y_t$-dependence is inherently two-dimensional.  While the two-dimensional dependence is interesting in its own right, we construct a one-dimensional function on $y_t$ to compare to published $v_2 ( p_t )$ data.

We have made nine marginal $y_t$ cuts---as described in Sec.~\ref{correlations}---for each centrality class of events for both 62 and 200 GeV collisions.  The first bin includes particles from 0.15~GeV---the lowest $p_t$ we can measure at STAR---to a $y_t$ value of 1.4.  Above that there are 7 bins evenly spaced in $y_t$ with a width of 0.4 units of transverse rapidity.  The final bin includes all particles greater than $y_t = 4.2$.  In $p_t$ these bin edges correspond to
0.27~GeV, 0.41~GeV, 0.62~GeV, 0.94~GeV, 1.4~GeV, 2.1~GeV, 3.1~GeV, and 4.7~GeV.

\begin{figure*}
\resizebox{1.00\textwidth}{!}{
  \includegraphics{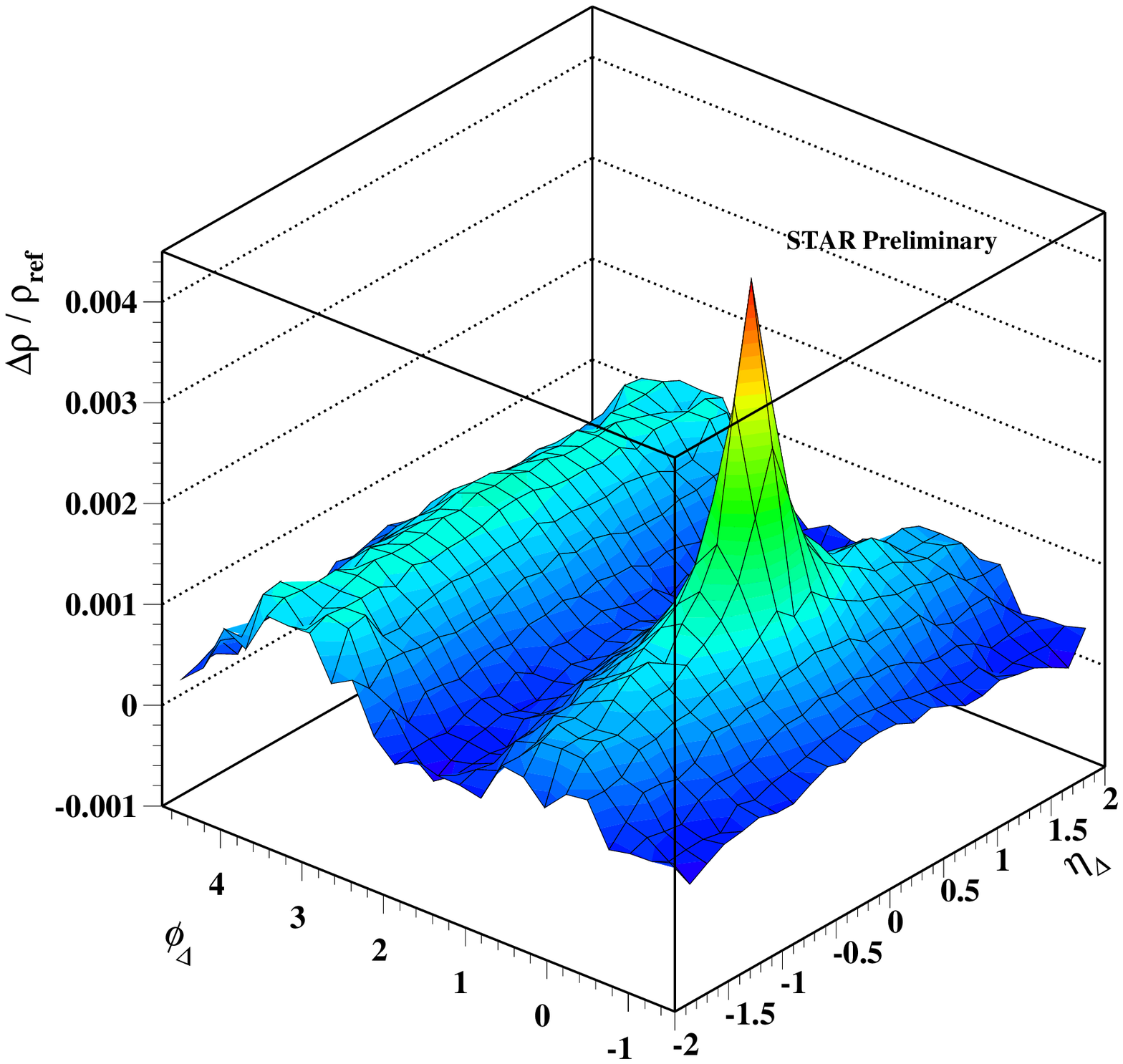}
  \includegraphics{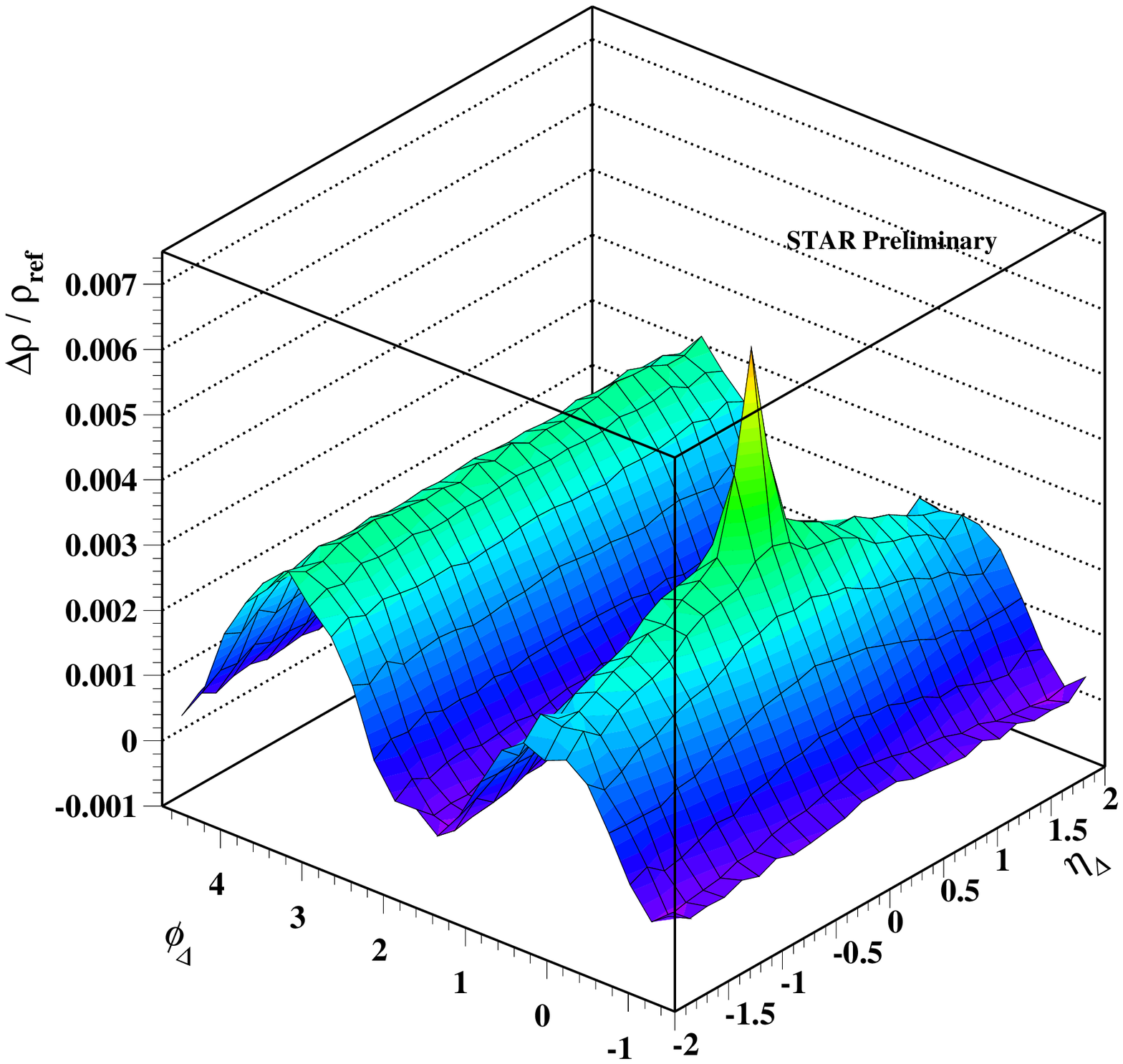}
  \includegraphics{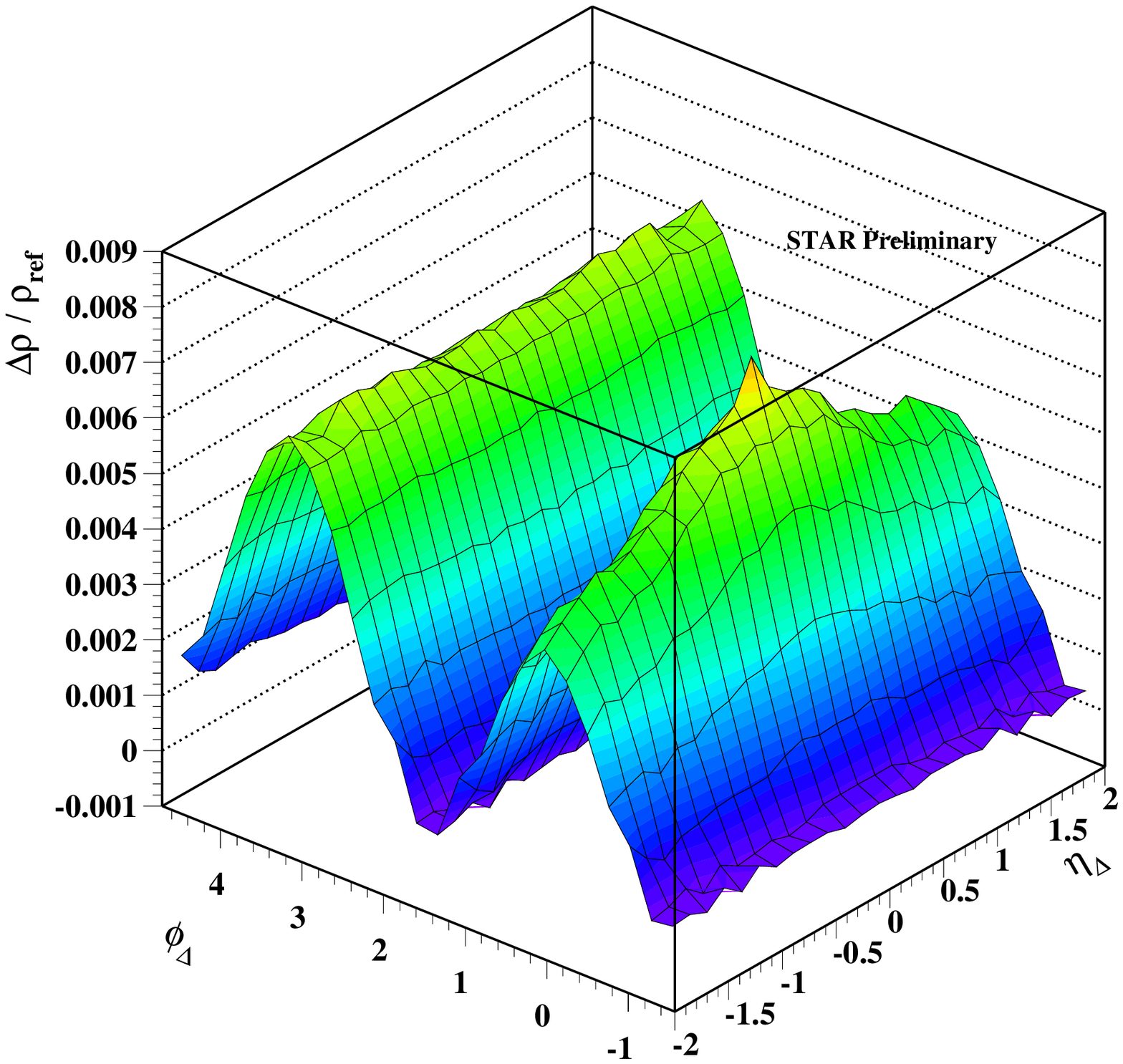}
  \includegraphics{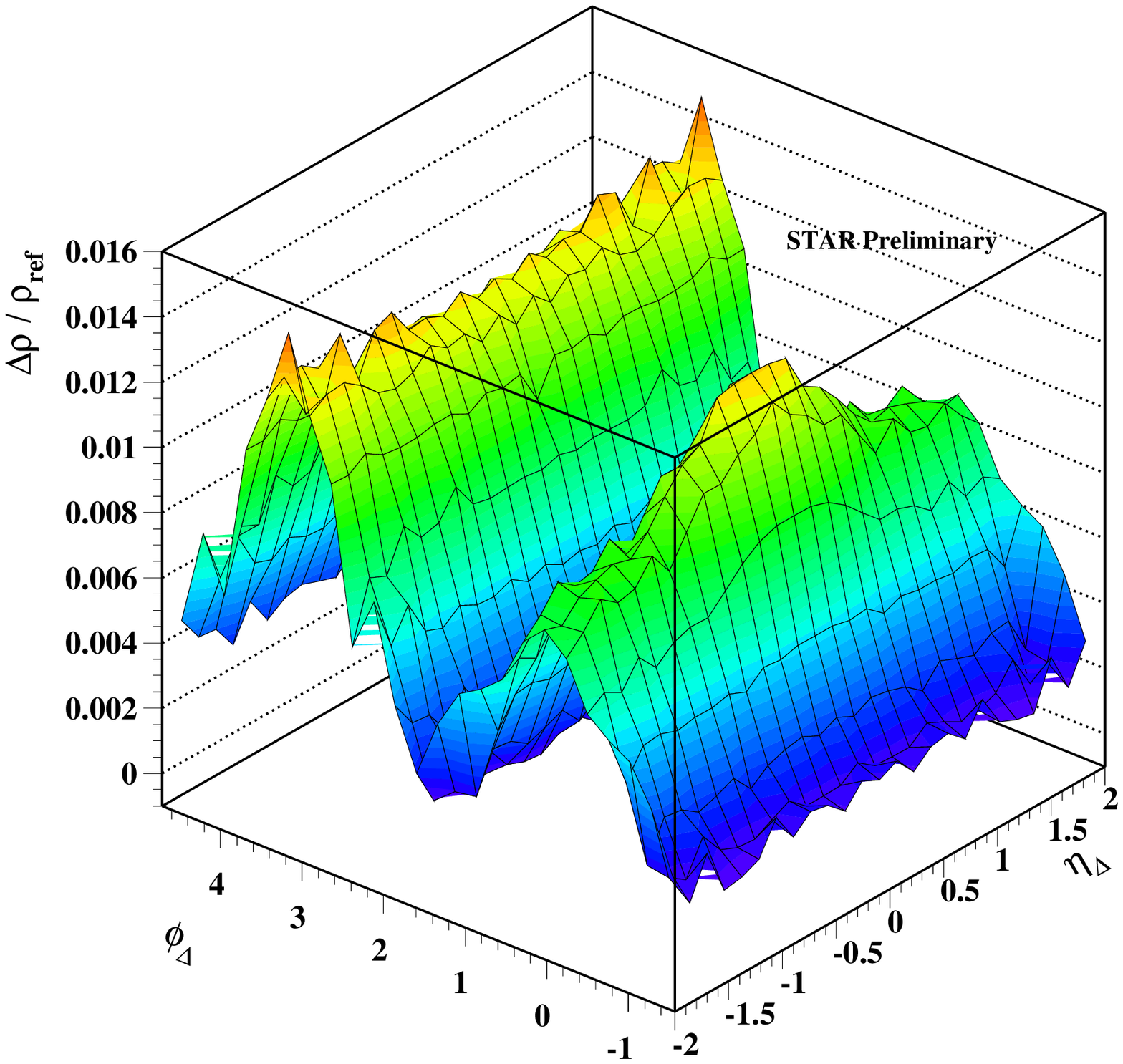}
}
\caption{\label{fig:examples62}Example of the $y_t$ evolution of correlation structures for 62 GeV 40-50\% central collisions.  The plots correspond to $y_t$ bins of $1.4 < y_t < 1.8$, $3.0 < y_t < 3.4$, and $3.8 < y_t
< 4.2$.}
\end{figure*}

\begin{figure*}
\resizebox{1.00\textwidth}{!}{
  \includegraphics{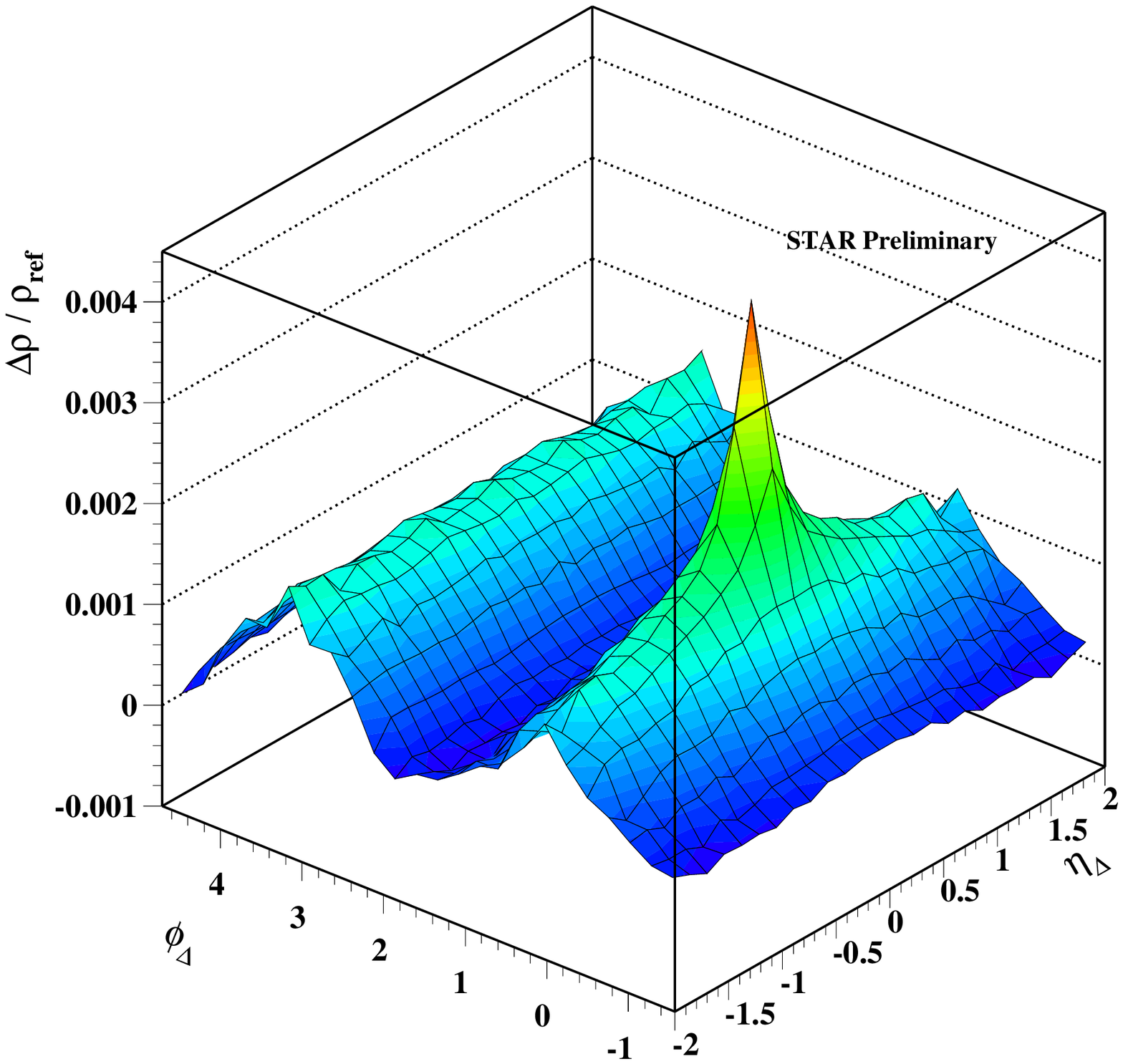}
  \includegraphics{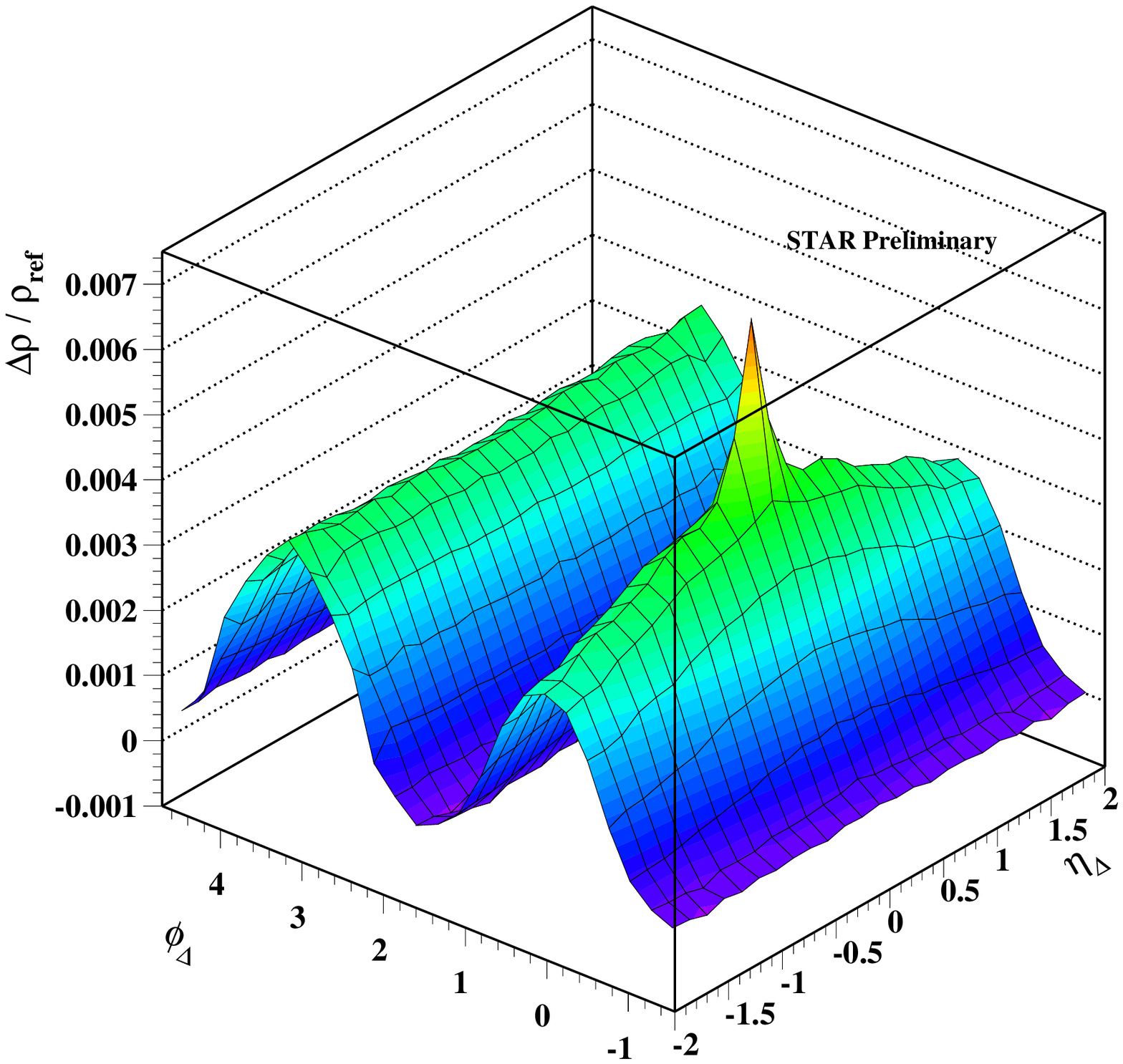}
  \includegraphics{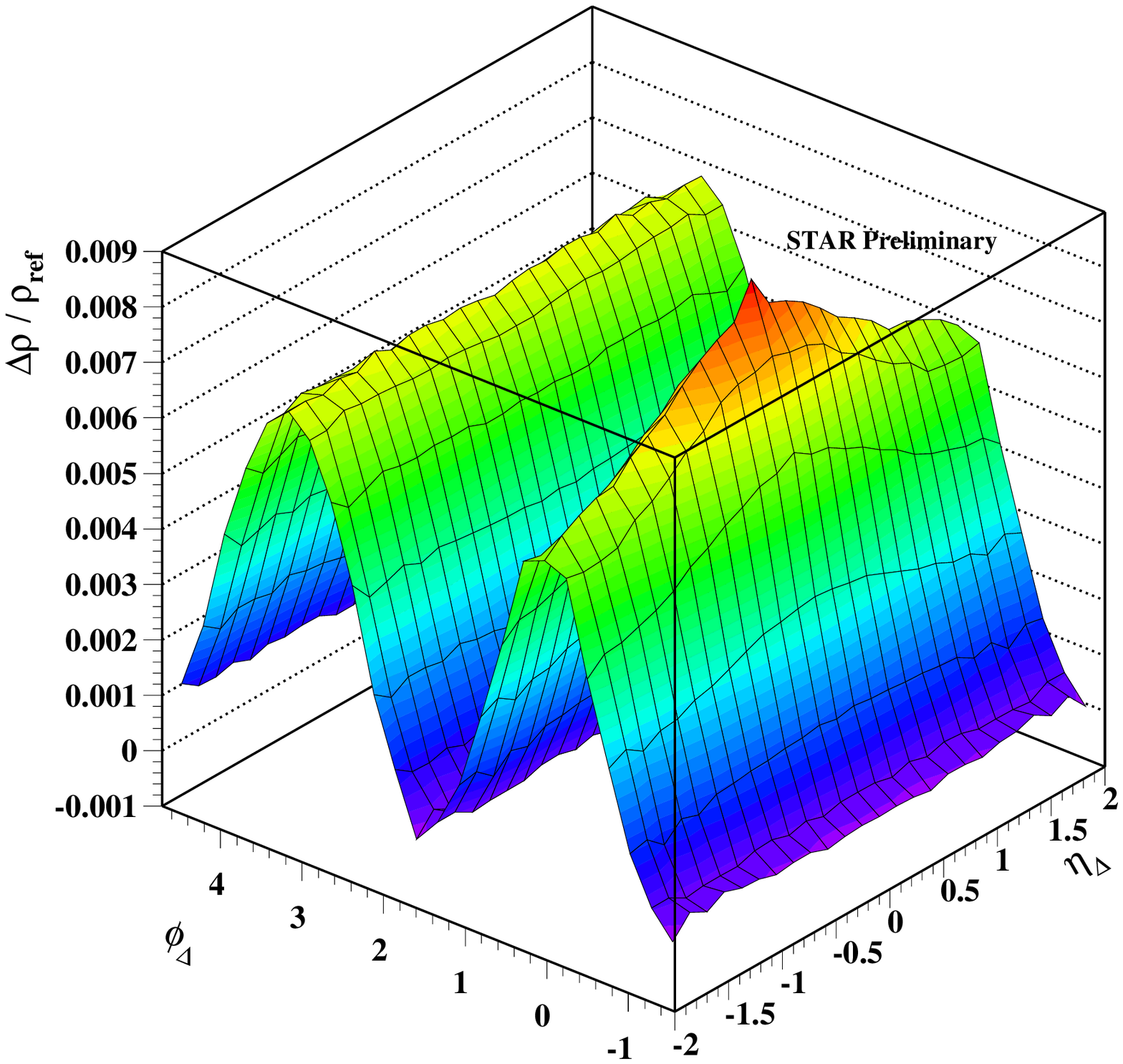}
  \includegraphics{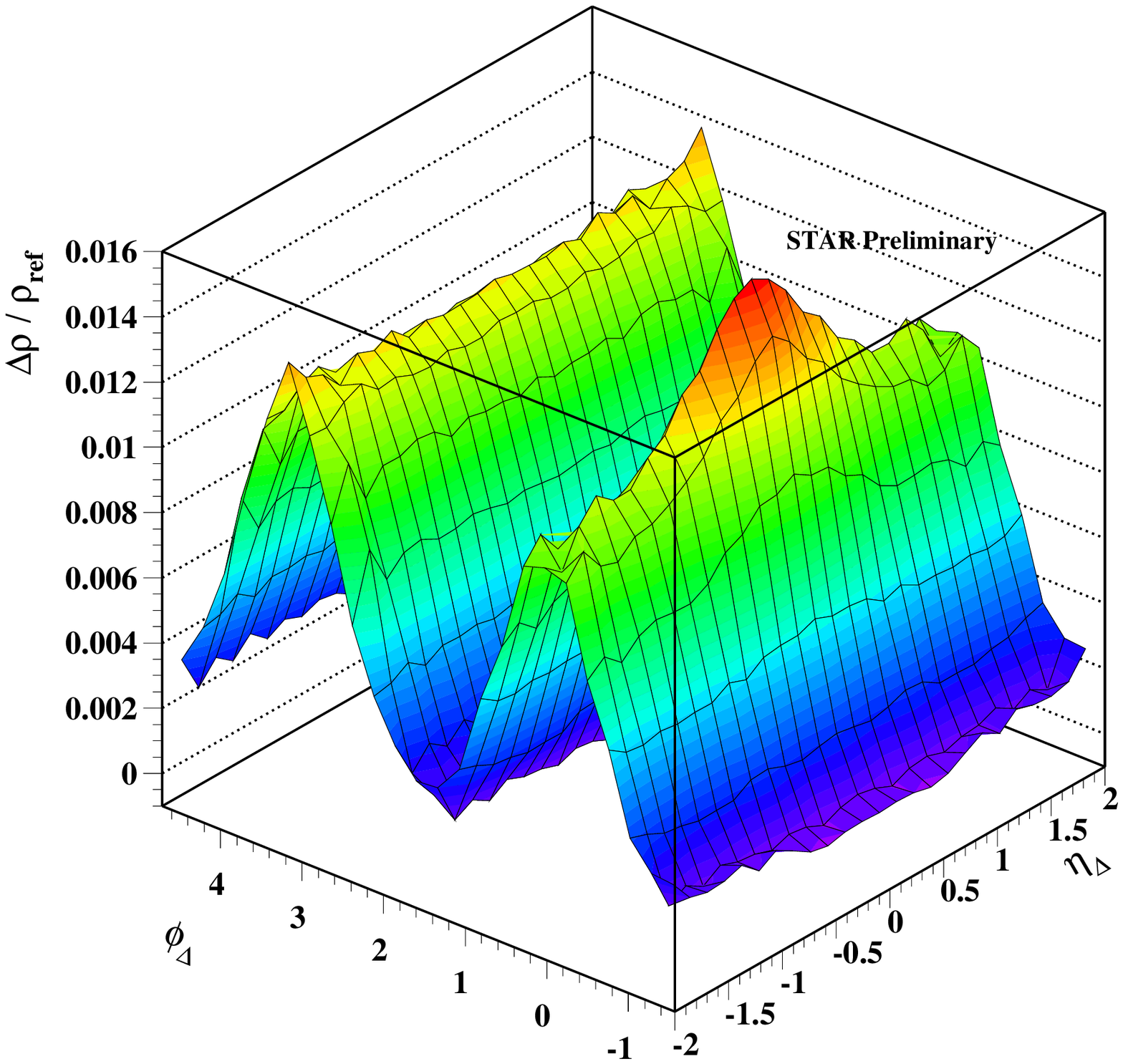}
}
\caption{\label{fig:examples200}Example of the $y_t$ evolution of correlation structures for 200 GeV 40-50\% central collisions.  The plots correspond to $y_t$ bins of $1.4 < y_t < 1.8$, $3.0 < y_t < 3.4$, and $3.8 < y_t < 4.2$.}
\end{figure*}

In Figs. \ref{fig:examples62} and \ref{fig:examples200} we show example histograms for 62 GeV and 200 GeV collisions respectively at 40-50\% centrality.  All of the $p_t$-dependent correlations are measured using $\Delta \rho / \rho_{\textrm{ref}}$ which simplifies the conversion to $v_2$.

The fit procedure is similar to that used in the $p_t$-integrated results with one major exception:  We no longer attempt to model the sharp exponential peak.  Instead we simply exclude bins near the angular origin from the fit.
The excluded bins are at $\eta_\Delta = 0$ and $\phi_\Delta = 0, \pm \pi / 12$ and at $\phi_\Delta = 0$ and $\eta_\Delta = \pm 0.08 , \pm 0.16$.
This has the effect of making the fits more stable over a wide $p_t$ range but it can be a problem for the most peripheral bins where the width of the exponential approaches that of the same-side 2D Gaussian.

\begin{figure*}
\resizebox{1.00\textwidth}{!}{
  \includegraphics{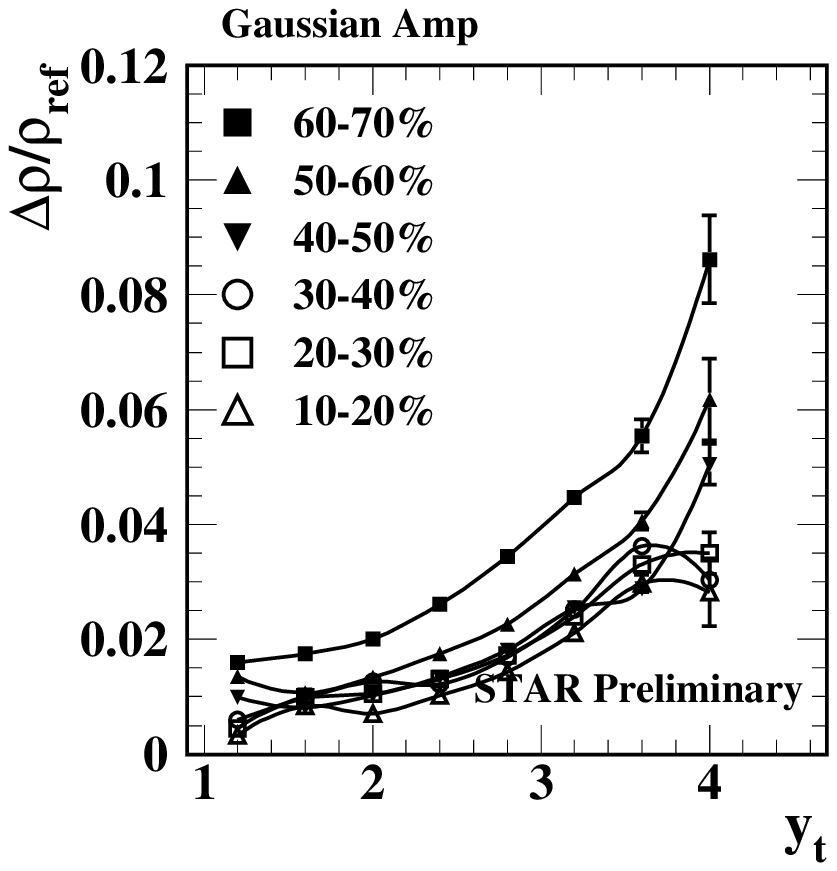}
  \includegraphics{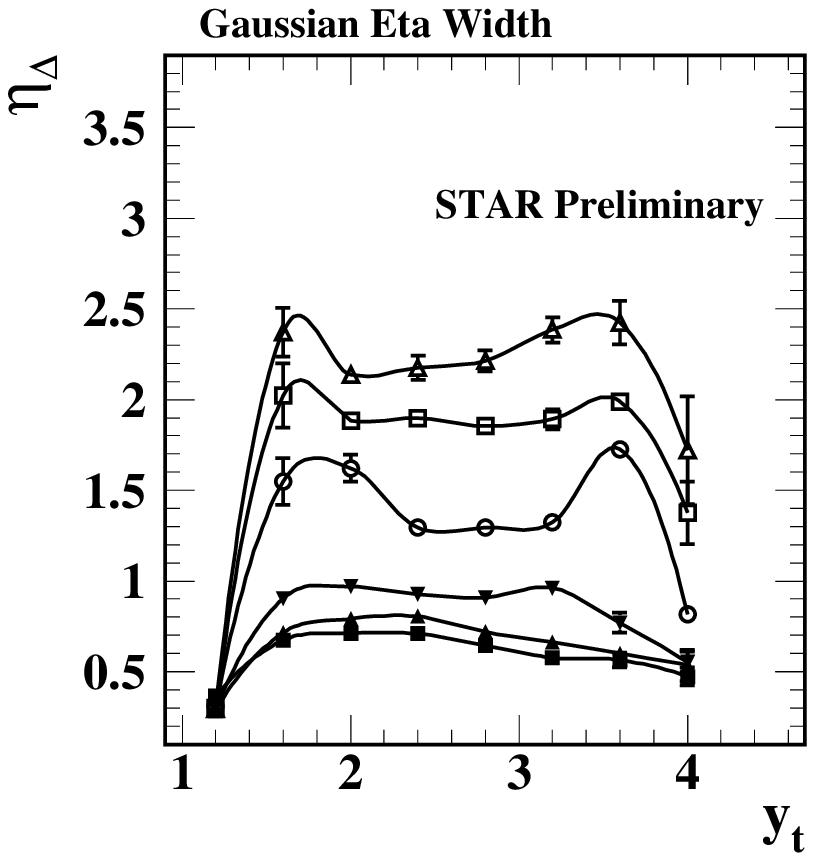}
  \includegraphics{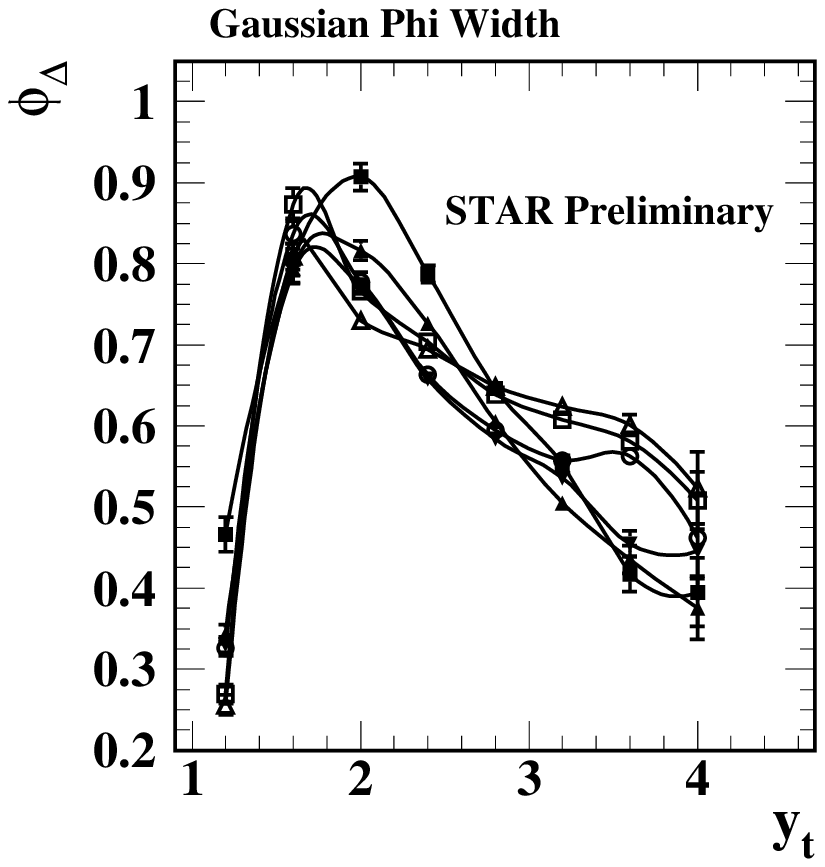}
  \includegraphics{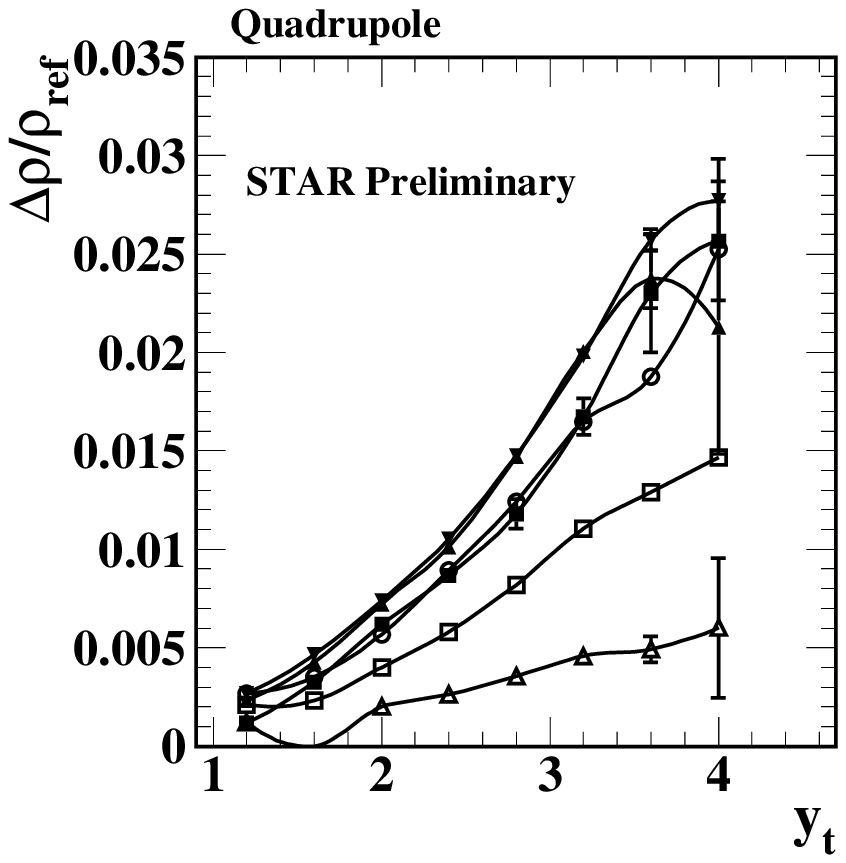}
}
\caption{\label{fig:fit62}Fit parameters for 62 GeV collisions: Same-side peak amplitude, width in eta, width in phi, and quadrupole amplitude as a function of $y_t$ for 10-20\%, 20-30\%, 30-40\%, 40-50\%, 50-60\%, and 60-70\% central collisions.  Error bars are for fitting errors only.}
\end{figure*}

\begin{figure*}
\resizebox{1.00\textwidth}{!}{
  \includegraphics{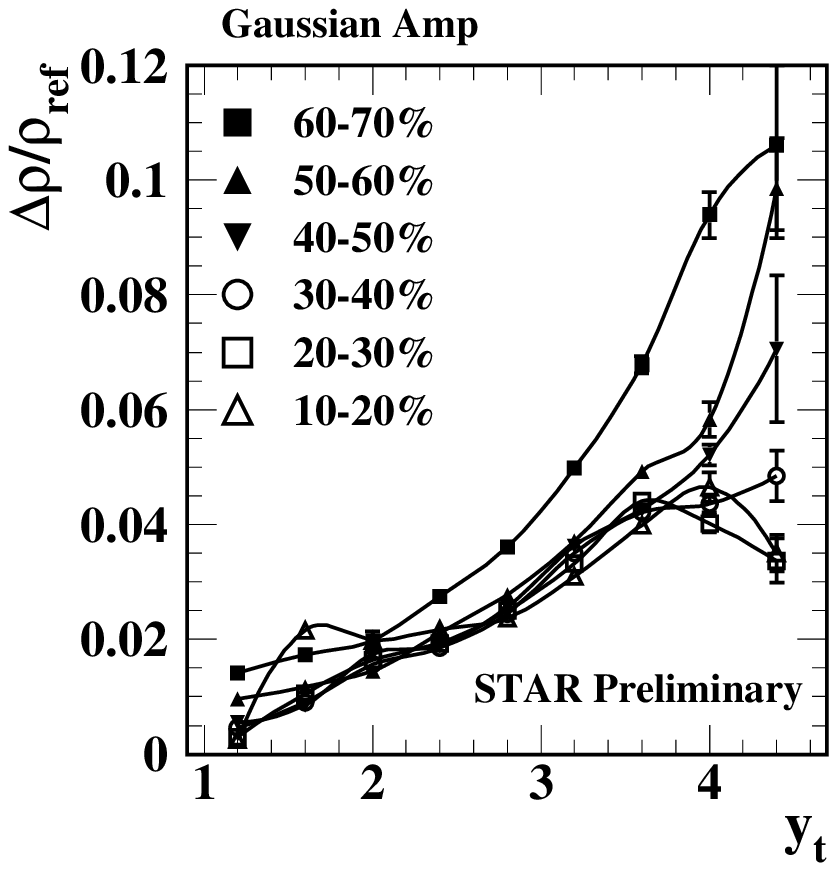}
  \includegraphics{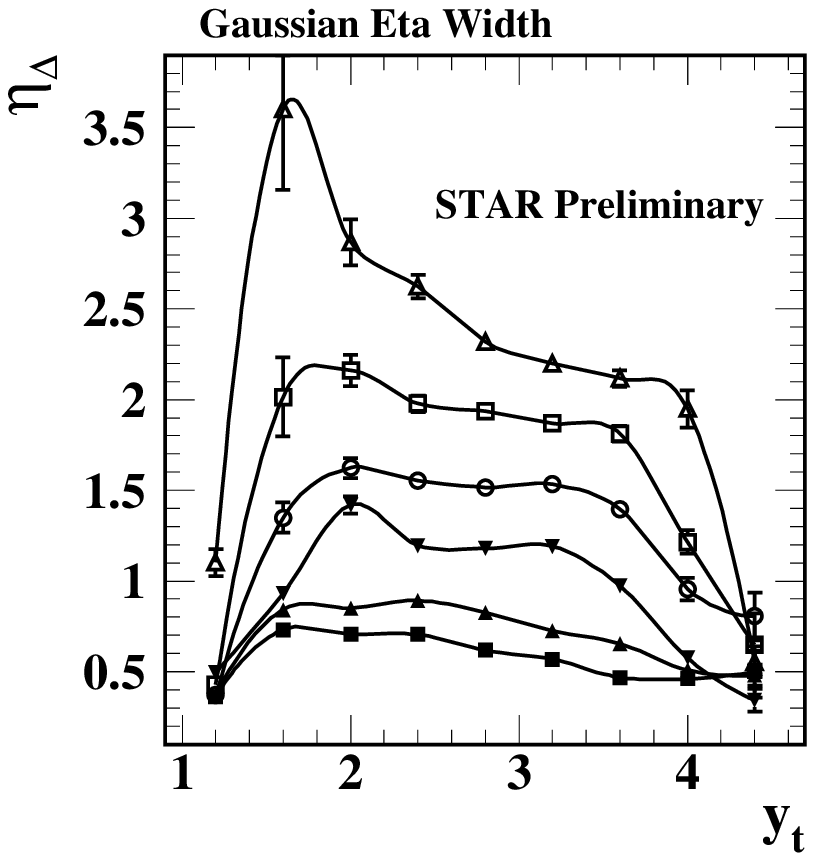}
  \includegraphics{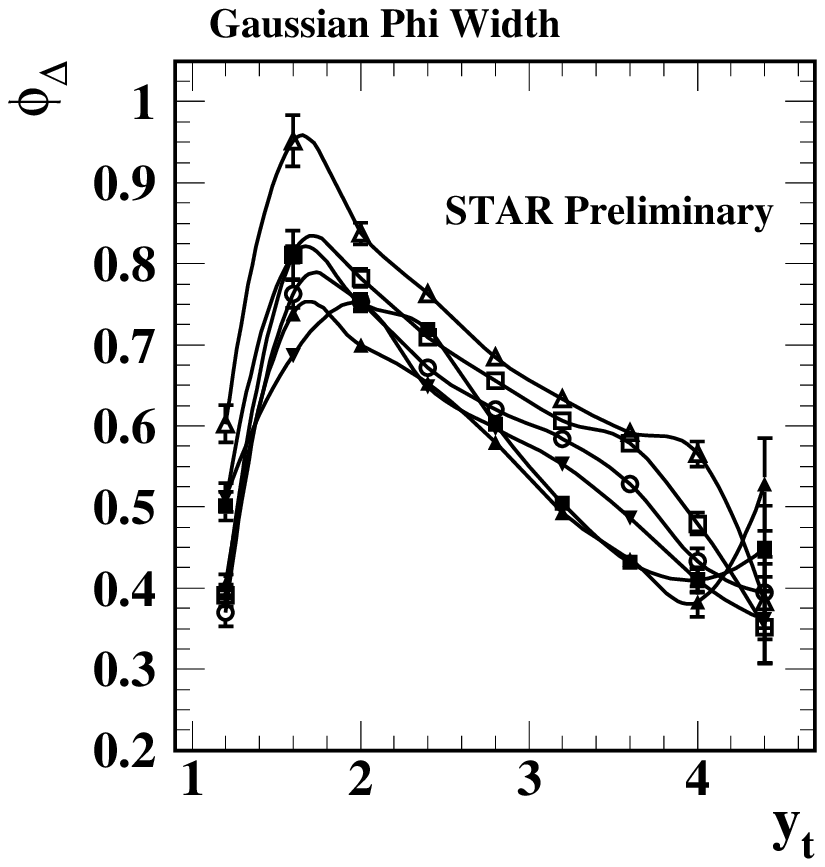}
  \includegraphics{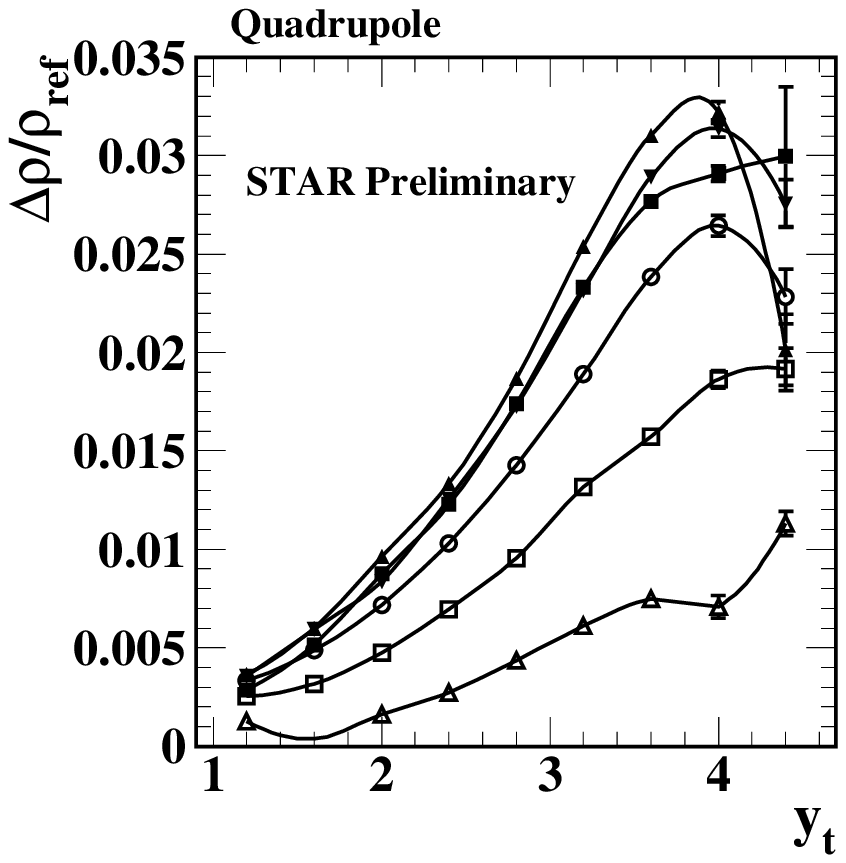}
}
\caption{\label{fig:fit200}Fit parameters for 200 GeV collisions: Same-side peak amplitude, width in eta, width in phi, and quadrupole amplitude as a function of $y_t$ for 10-20\%, 20-30\%, 30-40\%, 40-50\%, 50-60\%, and 60-70\% central collisions.  Error bars are for fitting errors only.}
\end{figure*}

The major fit parameters are shown in Fig.~\ref{fig:fit62} for 62 GeV collisions and in Fig.~\ref{fig:fit200} for 200 GeV collisions for several centralities.  The Gaussian amplitudes and azimuth widths follow expected trends for both energies.  However, the widths on pseudorapidity seem to be largely independent of $y_t$ over a large range.

Quadrupole amplitudes in $\Delta \rho / \rho_{\textrm{ref}}$ can be converted into $v_2$ values by the simple relationship, $2 v_{2}^{2} \{ 2D \} \equiv \Delta \rho [2] / \rho_{\textrm{ref}}$.  However, the marginal distribution produces data of the form $2 v_{2} ( p_t , b) v_{2} (b)$, not  $2 v_{2}^{2} ( p_t , b)$, so the fit amplitude must be divided by $p_t$-integrated $v_2 (b)$.

\begin{figure*}
  \hspace{35pt}
\resizebox{.75\textwidth}{!}{
  \includegraphics{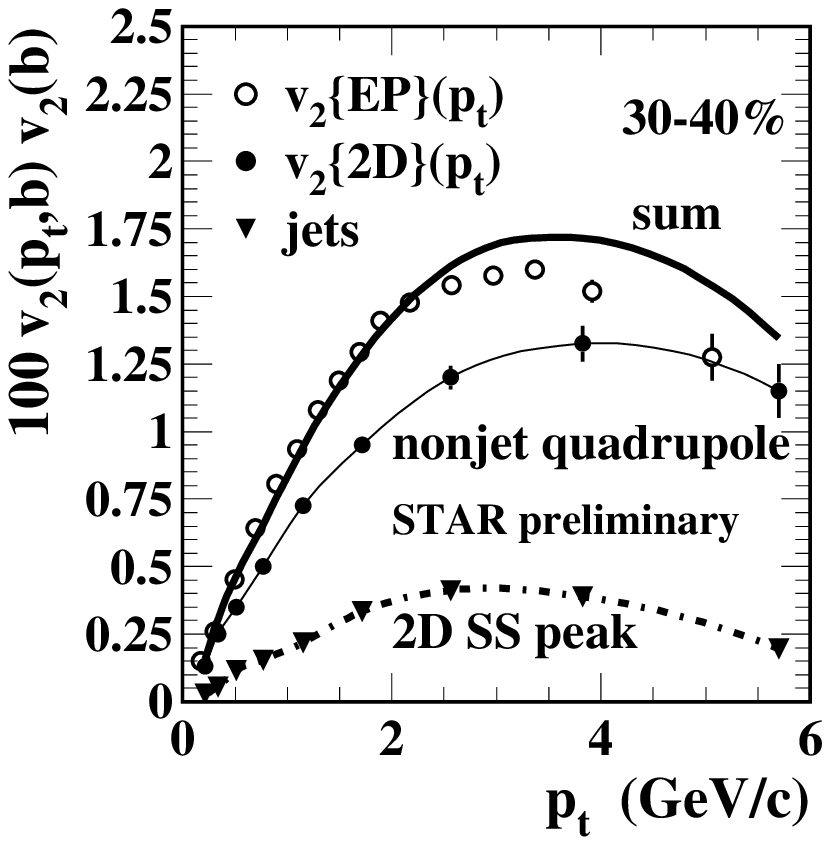}
  \includegraphics{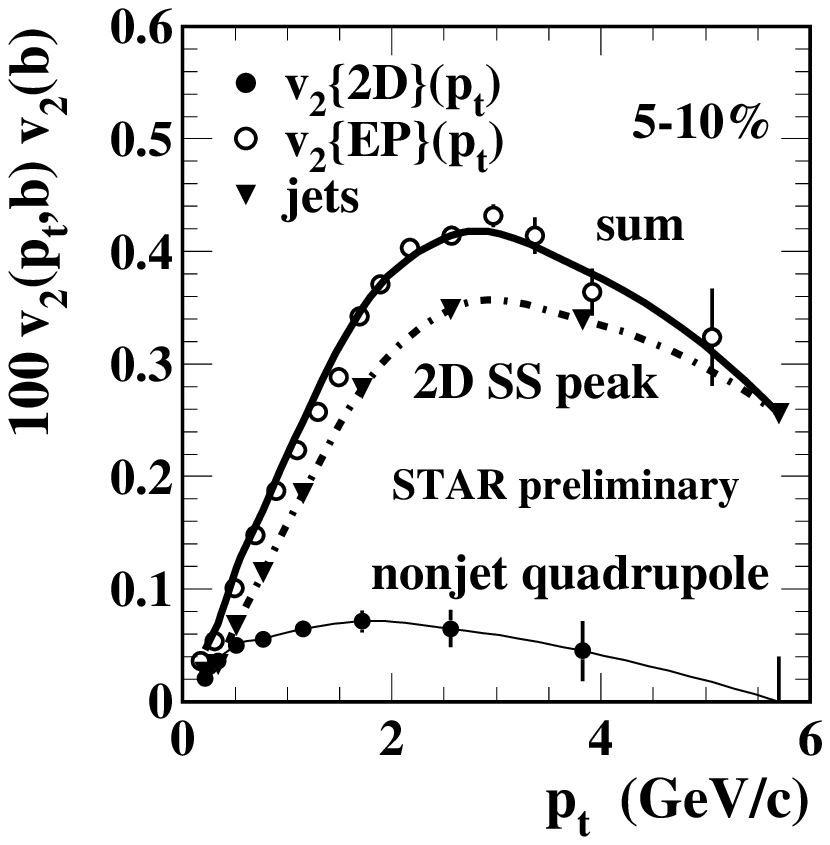}
  \includegraphics{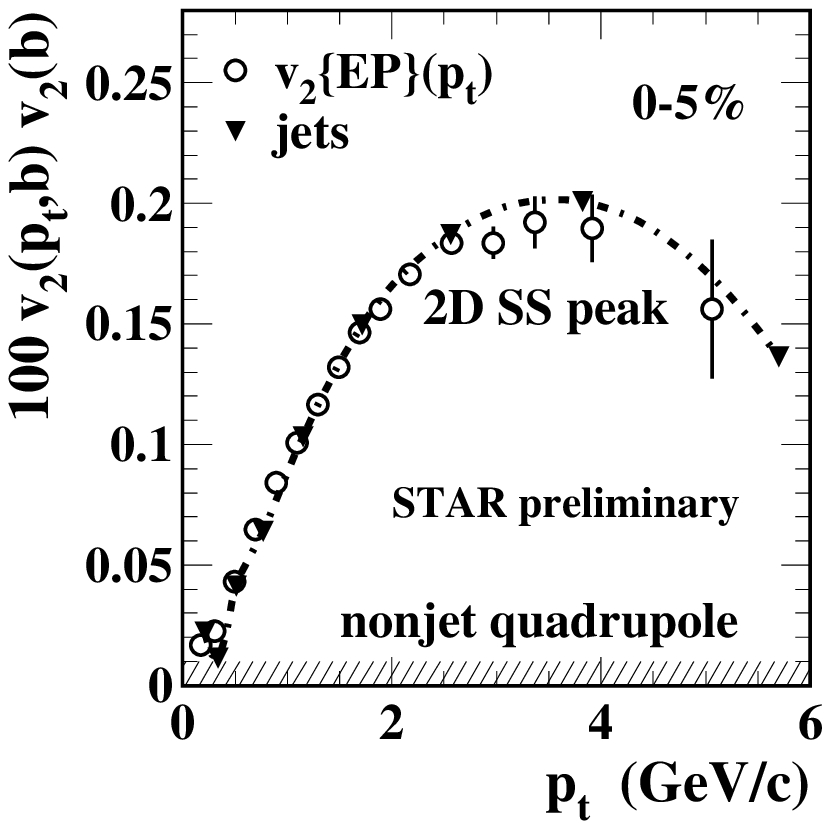}
}
\caption{\label{fig:v2EPcomp} Comparison of quadrupole (closed circles) and the second Fourier component of the 2D same-side Gaussian (closed upside-down triangles) to $v_2 \{ EP \}$ results (open circles) \cite{v2EP} for 30-40\%, 5-10\%, and 0-5\% central collisions.  The dark solid curve in the first two panels is the sum of the quadrupole and same-side peak terms.}
\end{figure*}

In our fit model the non-quadrupole term with the largest contribution to the second Fourier component on azimuth is the same-side 2D Gaussian peak.  We can calculate the contribution to the second Fourier component of a 2D Gaussian of given amplitude and widths, which we know for the same-side peak from our fit parameters.  This gives us a direct measure of the so-called ``nonflow'' contribution to $v_2$, as shown in Fig.~\ref{fig:v2EPcomp} for 30-40\%, 5-10\%, and 0-5\% central collisions.  We see that the 2D Gaussian peak dominates in 0-5\% central collisions and that the ``nonflow'' (jet contribution) is strongly centrality and $p_t$-dependent.

\section{Quadrupole Spectrum}

\begin{figure*}
  \hspace{35pt}
\resizebox{0.75\textwidth}{!}{
  \includegraphics{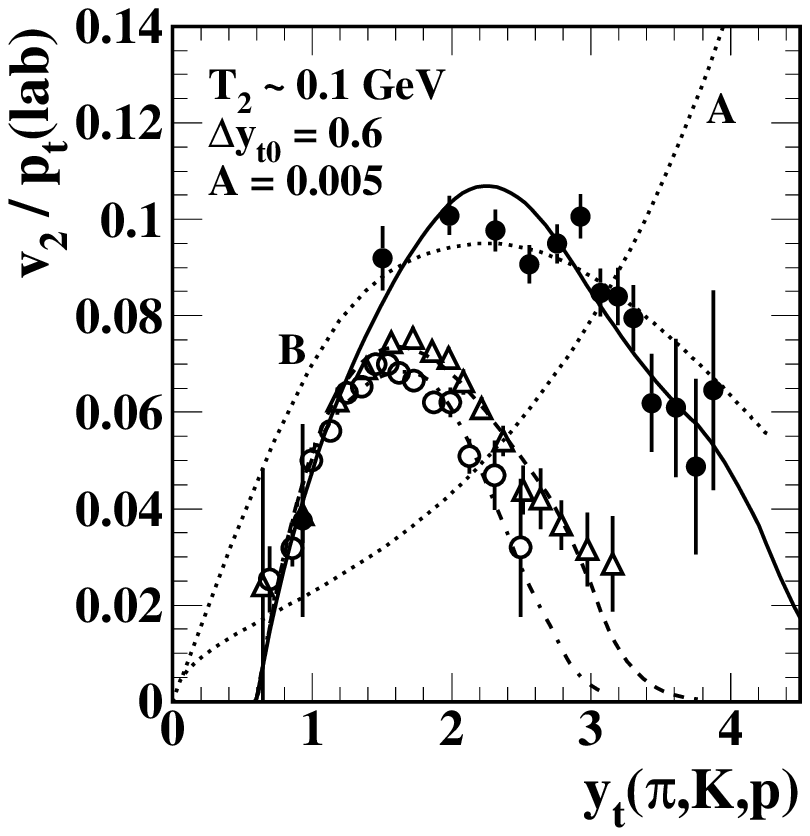}
  \includegraphics{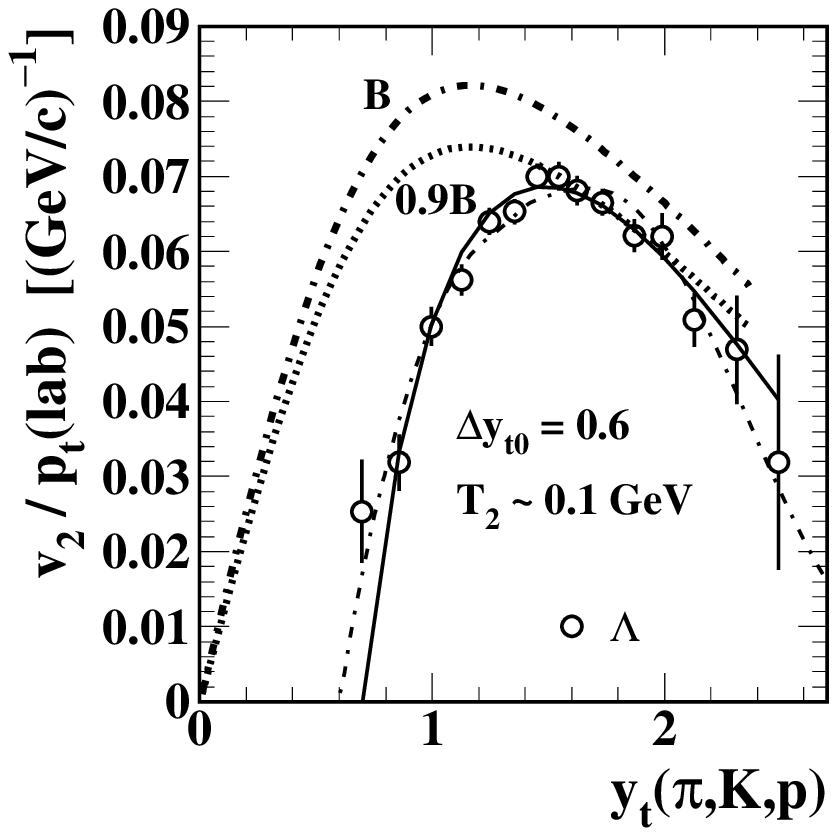}
  \includegraphics{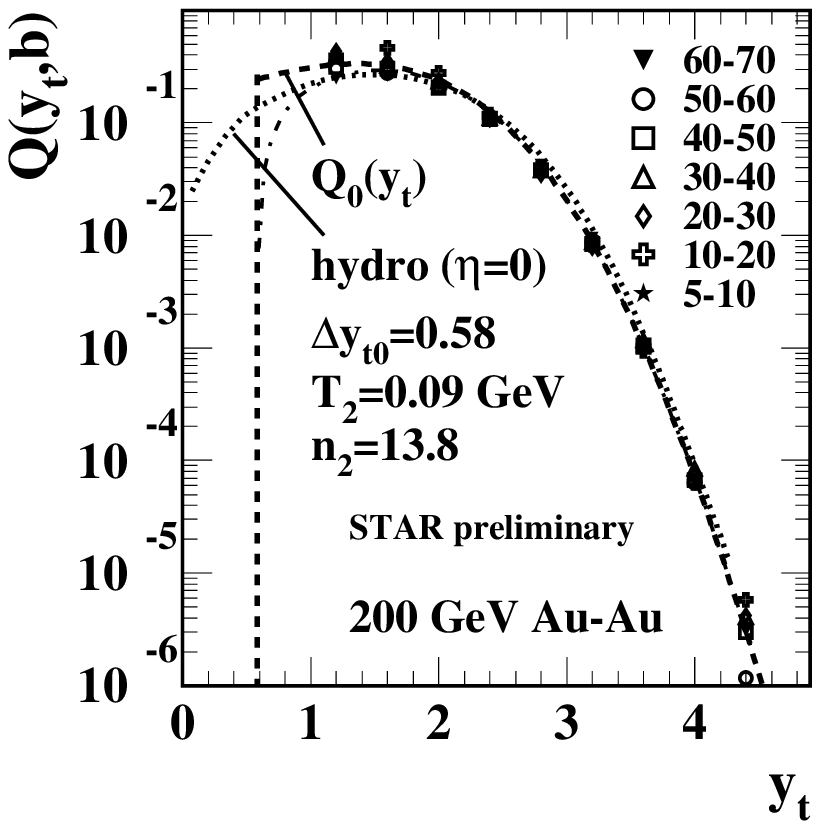}
}
\caption{\label{fig:quadspec}Left panel: Identified $v_2 (p_t) / p_t$ for pions (closed circles), kaons (open triangles), and protons (open circles) for minimum-bias collisions \cite{identifiedv2} vs. proper $y_t$ for the particle species.  Middle panel:  A magnification of the proton results.  Right panel: The quadrupole spectrum for a range of centralities at 200~GeV.}
\end{figure*}

Consider the case of minimum-bias identified particles results from STAR \cite{identifiedv2}.  In Ref.~\cite{quadspec} it was shown that by plotting $v_2 / p_t$ on proper $y_t$ for identified hadron species the particles appear to come from a single boosted source.  This is reproduced in Fig.~\ref{fig:quadspec} (left panel).  This is most evident for the more massive particles.  Since the pion mass is of the same order as the lowest $p_t$ we can measure in STAR there are few observed pions in the necessary $p_t$ range to see this boost.  In Fig.~\ref{fig:quadspec} (middle panel) we plot just the Lambdas, which makes the boost more explicit.

Now consider the centrality-dependent (denoted by $b$) production of an azimuth quadrupole term from a general boosted source.  First make the general assumption that the single-particle density on $y_t$ and $\phi$ can be decomposed into azimuth-dependent and azimuth-independent terms:
\begin{eqnarray}
\rho (y_t , \phi , b) = \rho_0 (y_t , b) + \rho_2 (y_t , \phi , b) .
\label{eqn:densities}
\end{eqnarray}
At this point we do not make any assumptions about the nature of the azimuth dependence.
Then apply the continuum definition of $v_2$:
\begin{eqnarray}
v_2 (y_t , b ) \hspace{-.05in} \equiv \hspace{-.05in} \frac{\frac{1}{2\pi} \int_{0}^{2\pi} \hspace{-.05in} d\phi \rho (y_t
, \phi , b) \cos [2(\phi \hspace{-.03in} - \hspace{-.03in} \psi_R )]}{\frac{1}{2\pi} \int_{0}^{2\pi} d\phi \rho (y_t , \phi , b)}
\hspace{-.05in} \approx \hspace{-.05in} \frac{\frac{1}{2\pi} \int_{0}^{2\pi} \hspace{-.05in} d\phi \rho_2 (y_t , \phi , b) \cos [2(\phi \hspace{-.03in} - \hspace{-.03in} \psi_R )]}{\rho_0 (y_t , b)}
\hspace{-.05in} \equiv \hspace{-.05in} \frac{V_2}{\rho_0 ( y_t , b )} .
\label{eqn:v2def}
\end{eqnarray}
Only the azimuth-dependent term contributes to the numerator, and the denominator is approximately the single-particle spectrum \cite{quadspec}.  What we want to study is actually the numerator of this expression, which we denote by $V_2$.

To calculate $V_2$ we need to introduce a boost model.  A general boost in nuclear collisions should have both monopole (radial flow, Hubble expansion) and quadrupole terms, which is easily expressed in $y_t$:
\begin{eqnarray}
\Delta y_t ( \phi ) = \Delta y_{t0} + \Delta y_{t2} \cos (2 [ \phi - \psi_R ] ) ,
\label{eqn:dytdef}
\end{eqnarray}
with $\Delta y_{t2} \leq \Delta y_{t0}$ a necessary condition for a positive-definite boost.  Using a simple blast-wave model with a Maxwell-Boltzmann distribution for a locally-thermalized source the boosted spectrum's azimuth-dependent term then has the form \cite{quadspec}:
\begin{eqnarray}
\rho_2 ( y_t , \phi ) = A_{2y_{t}} \exp \{ - \mu_2 [ \cosh ( y_t - \Delta y_t ( \phi ) ) - 1 ] \} ,
\label{eqn:boostedspec}
\end{eqnarray}
where $\mu_2 = m_0 / T_2$.
If we insert our boost model into Eq.~\ref{eqn:boostedspec} and factor the $\phi$-dependent terms of $\rho_2$ into the form $\rho_2 ( y_t , \phi ) = \rho_2 ( y_t ) \times F_1 ( y_t , \phi ) \times F_2 ( y_t , \phi )$ where
\begin{eqnarray}
F_1 ( y_t , \phi ) &=& \exp \{ m_{t}^{\prime} [ \cosh ( \Delta y_{t2} \cos [ 2 ( \phi - \psi_r ) ] ) - 1 ] / T_2 \} \nonumber \\
F_2 ( y_t , \phi ) &=& \exp \{ p_{t}^{\prime} [ \sinh ( \Delta y_{t2} \cos [ 2 ( \phi - \psi_r ) ] ) / T_2 \}
\label{eqn:Fdefs}
\end{eqnarray}
then we identify $\rho_2 ( y_t )$ as the azimuth-integrated quadrupole spectrum. 
Inserting this into the numerator of Eq.~\ref{eqn:v2def} and evaluating the integral yields
\begin{eqnarray}
V_2 (y_t , b) \approx \frac{p_t \Delta y_{t2} (b)}{2 T_2} \rho_2 (y_t , b) .
\label{eqn:blastwave}
\end{eqnarray}

Finally, we want to isolate $\rho_2 (y_t , b)$ by taking the unit-integral ratio of measured quantities:
\begin{eqnarray}
Q(y_t , b) \equiv \frac{V_2 (y_t , b)/p_t }{V_2 (b) \left< 1/p_t \right>} \approx \frac{\rho_2 (y_t , b)}{\rho_2 (b)} .
\label{eqn:Qdef}
\end{eqnarray}
The parameters from the blast-wave model present in Eq.~\ref{eqn:blastwave} then drop out in the ratio and $Q(y_t , b)$, shown in Fig.~\ref{fig:quadspec} (right panel), directly relates measured parameters to the quadrupole spectrum for 5-10\%, 10-20\%, 20-30\%, 30-40\%, 40-50\%, 50-60\%, and 60-70\% central 200~GeV collisions.

We observe approximate centrality-independence of $Q ( y_t , b )$ in Fig.~\ref{fig:quadspec} (right panel).  We conjecture that there is a centrality-independent $Q_0 ( y_t )$ which is well-described by a boosted L\'evy distribution with parameters $T_2 = 0.09 \textrm{ GeV}$, $n_2 = 13.8$, and $\Delta y_{t0} = 0.58$ (dashed curve), though there are significant deviations for the more-central events at low $p_t$.  This boost is the same that was observed in the case of minimum-bias identified hadrons in Fig.~\ref{fig:quadspec} (left and middle panels).



\section{Quadrupole Parametrization}

A detailed study of the systematics of the $p_t$-dependent quadrupole is possible.  First, consider rearranging the definition of $Q$ in Eq.~\ref{eqn:Qdef} and use the definition of $V_2$ from Eq.~\ref{eqn:v2def} to get
\begin{eqnarray}
v_2 \{ 2D \} ( p_t , b) = \left< \frac{1}{p_t} \right> p_t v_2 \{ 2D \} ( b ) \left[ \frac{\rho_0 ( b ) Q_0 ( p_t )}{\rho_0 ( p_t , b )} \right] .
\label{eqn:paramMotivate}
\end{eqnarray}
The quantity $\rho_0 ( b ) Q_0 ( p_t ) / \rho_0 ( p_t , b )$ has a $p_t$ dependence described by the ratio of a L\'evy distribution to the single-particle spectrum.  This is observed to be exponential for larger values of $p_t$.

We can then construct a new parametrization of the form
\begin{eqnarray}
v_2 \{ 2D \} ( p_t , b) \approx \left< \frac{1}{p_t} \right> p_t v_2 \{ 2D \} ( b ) \exp ( - p_t / 4 ) \times f ( p_t , b),
\label{eqn:paramatrization}
\end{eqnarray}
where $f ( p_t , b )$ is a dimensionless factor needed to describe deviations from the exponential form at low-$p_t$.  It can be fit to the data with the form
\begin{eqnarray}
f ( p_t , b ) = 1 + C ( b ) [ \textrm{erf} ( y_t - 1.2 ) - \textrm{erf} (1.8 - 1.2) ],
\label{eqn:ffunction}
\end{eqnarray}
where  $C(b) = 0.12 - (\nu - 3.4 ) / 5 - [ ( \nu - 3.4 ) / 2 ]^5$.  This parametrization provides a more accurate description of the quadrupole term over a wider range of $p_t$ and centrality than the L\'evy distribution.  The factor $f ( p_t , b )$ is approximately 1 above about $0.75$~GeV/c.  Above that point the $p_t$ dependence of the quadrupole is entirely described by the factor $p_t \exp ( - p_t / 4 )$.  This leads to a factorization of the $p_t$ and centrality dependence of $v_2 ( p_t , b )$ for these higher $p_t$s.

This factorization can be combined with the factorization of collision energy and centrality dependence of the $p_t$-integrated quadrupole described in Sec.~\ref{integrated} for a complete description of the azimuth quadrupole component, at least at higher $p_t$.  This implies that there is very simple underlying behavior of the azimuth quadrupole.

\section{Conclusions}


Two-particle correlation histograms on azimuth and pseudorapidity have been constructed for a wide range of centrality and momentum conditions.  These histograms can be fit to study different aspects of the physical system.  The quadrupole term is closely related to the standard definition of $v_2$ but is isolated from $\eta$-dependent ``nonflow'' effects.

Studies of the $p_t$-integrated quadrupole \cite{quadrupole} have revealed simple trends on collision energy and centrality.  The dependence on collisions energy and centrality can be factorized to produce a very accurate description of $p_t$-integrated data.

We construct $p_t$-dependent histograms using marginal distributions.  Published event-plane $v_2 ( p_t , b)$ data \cite{v2EP} are accurately described by the sum of the quadrupole and 2D Gaussian fit components.
The $p_t$-dependent quadrupole component is used to construct a boosted quadrupole spectrum.  The quadrupole spectrum is approximately centrality-independent and is well described by a fixed boosted L\'evy distribution.  An accurate parametrization of the $p_t$-differential quadrupole exhibits simple scaling above $0.75$~GeV/c.

Quadrupole systematics reveal a system with remarkably simple scaling behavior and a possible factorization of the collision energy, centrality, and $p_t$ dependence.  This seems contrary to typical hydrodynamic expectations in nuclear collisions.


\begin{thebibliography}{99}
\bibitem{starreview}
J.\ Adams \emph{et al.} (STAR Collaboration), Nucl.\ Phys.\ A \textbf{757}, 102 (2005).
\bibitem{hydrointro}
D.\ Teaney, J.\ Lauret, and E.V.\ Shuryak, Phys.\ Rev.\ Lett.\ \textbf{86}, 4783 (2001).
\bibitem{basicv2}
A.M.\ Poskanzer, S.A.\ Voloshin, Phys.\ Rev.\ C \textbf{58}, 1671 (1998).
\bibitem{azstruct}
T.A.\ Trainor and D.T.\ Kettler, Int.\ J.\ Mod.\ Phys.\ E \textbf{17}, 1219 (2008).
\bibitem{ZYAM}
T.A.\ Trainor, Phys.\ Rev.\ C \textbf{81}, 014905 (2010).
\bibitem{centralities}
T.A.\ Trainor and D.J.\ Prindle, hep-ph/0411217
\bibitem{quadrupole}
D.T.\ Kettler (STAR Collaboration), Eur.\ Phys.\ J.\ C \textbf{62}, 175 (2008).
\bibitem{mikeQM}
M.\ Daugherity (STAR Collaboration), J.\ Phys.\ G \textbf{35}, 104090 (2008).
\bibitem{fragevolve}
T.A.\ Trainor,
  Phys.\ Rev.\  C \textbf{80}, 044901 (2009).
\bibitem{v24}
C.\ Adler \emph{et al.} (STAR Collaboration), Phys.\ Rev.\ C \textbf{66}, 034904 (2002).
\bibitem{na49}
A.M.\ Poskanzer \emph{et al.} (NA49 Collaboration), Nucl.\ Phys.\ A \textbf{661}, 341 (1999).
\bibitem{ecc}
H. Sorge, Phys. Rev. Lett. \textbf{82}, 2048 (1999).
\bibitem{jaccoop}
P. Jacobs and G. Cooper, nucl-ex/0008015v1
\bibitem{part}
R. S. Bhalerao and J. Y. Ollitrault, Phys. Lett. B \textbf{641}, 260 (2006).
\bibitem{tcspectra}
T.A.\ Trainor, Int.\ J.\ Mod.\ Phys.\ E \textbf{17}, 1499 (2008).
\bibitem{v2EP}
J.\ Adams \emph{et al.} (STAR Collaboration), Phys.\ Rev.\ C \textbf{72}, 014904 (2005).
\bibitem{identifiedv2}
B.I.\ Abelev \emph{et al.} (STAR Collaboration), Phys.\ Rev.\ C \textbf{75}, 054906 (2007).
\bibitem{quadspec}
T.A.\ Trainor, Phys.\ Rev.\ C \textbf{78}, 064908 (2008).


\end{thebibliography}
\end{document}